\documentclass[twoside,nohyper]{JHEP3} 
\usepackage{epsfig,graphicx,dsfont}

\newcommand{\ba}{\begin{eqnarray}}
\newcommand{\ea}{\end{eqnarray}}
\newcommand{\be}{\begin{equation}}
\newcommand{\ee}{\end{equation}}
\newcommand{\dis}{\displaystyle}
\newcommand{\tr}{\mbox{tr}}
\newcommand{\re}{\mbox{Re }}
\newcommand{\im}{\mbox{Im }}
\newcommand{\order}{{\cal O}}

\newcommand{\mkd}{m_K^2}
\newcommand{\mpd}{m_{\pi}^2}

\newcommand{\eq}{\begin{eqnarray}}
\newcommand{\en}{\end{eqnarray}}

\newcommand{\et}{\!\!\!&&\!\!\!}
\newcommand{\fr}{\frac}
\newcommand{\al}{\alpha}
\newcommand{\la}{\lambda}
\newcommand{\de}{\delta}
\newcommand{\s}{\sigma}
\newcommand{\nn}{\nonumber}

\newcommand{\barr}{\begin{array}{c}}
\newcommand{\earr}{\end{array}}

\title{Charged Kaon $K \to 3\pi$ CP Violating Asymmetries\\ 
at NLO in CHPT} 
\author{Elvira G\'amiz and Joaquim Prades\\
Centro Andaluz de F\'\i sica de las Part\'\i culas Elementales 
(CAFPE) and \\ 
Departamento de F\'\i sica Te\'orica y del Cosmos, 
Universidad de Granada,\\ Campus de Fuente Nueva, E-18002 Granada, Spain}
\author{Ignazio Scimemi\\
 Institute of Theoretical Physics, University of Bern,\\ Sidlerstr. 5, 
CH-3012 Bern, Switzerland}

 \abstract{We give  the first full  
 next-to-leading order analytical 
results in Chiral Perturbation Theory for the charged
Kaon $K \to 3 \pi$
slope $g$ and decay rates CP-violating asymmetries.
We have included the dominant Final State Interactions
 at NLO analytically and discussed 
the importance of the unknown counterterms. 
We find that the uncertainty due to them is reasonable 
just for $\Delta g_C$, i.e.
the  asymmetry in the $K^+ \to \pi^+ \pi^+ \pi^-$ slope $g$,
 we get $\Delta g_C=-(2.4 \pm 1.2)\times 10^{-5}$.
 The rest of the asymmetries
are very sensitive to the unknown counterterms,
in particular,  the decay rate asymmetries can change even  sign.
One can use  this large sentivity to get  valuable information
on those counterterms and on $\im G_8$ coupling --very important 
for the CP-violating parameter $\varepsilon'_K$-- 
from the eventual measurement of these asymmetries.
We also provide the one-loop ${\cal O}(e^2 p^2)$ 
electroweak octet  contributions for the neutral and charged
Kaon  $K \to 3 \pi$ decays.}

\preprint{BUTP-2003-12\\CAFPE-19/03\\ UGFT-149/03\\hep-ph/0309172\\
October 2003\\Revised}

\begin{document}

\section{Introduction}

  The decay of a Kaon into three pions has a long history.
The first calculations were done 
 using current algebra methods or tree level Lagrangians,
see \cite{treelevel}  and references therein. 
Then using Chiral Perturbation Theory
(CHPT) \cite{WEI79,GL} at tree level in \cite{CHPTtree}.
Some introductory lectures on CHPT can be 
found in \cite{schools} and  recent reviews in \cite{reviews}.

The one-loop calculation was done in \cite{KMW91,KMW90} and used
in \cite{KDHMW92}, unfortunately
the analytical full results were not available. Recently,  there has 
appeared the first full published result in \cite{BDP03}.

CP-violating observables in $K \to 3 \pi$ decays 
have also attracted a lot of work since long time ago 
\cite{HOL69,LW80,AVI81,GRW86,DHV87,BBEL89,DIP91,IMP92,MP95,SHA03,SHA93,DIPP94} and references therein.

At next-to-leading order (NLO)
 there were no exact results available in CHPT 
so that the results presented in \cite{BBEL89,DIP91,IMP92,MP95}
 about the NLO corrections were based in assumptions about  the behavior 
of those corrections and/or 
 using model depending results  in \cite{BBEL89}. 
In \cite{SHA03,SHA93} there are partial
 results at NLO within the linear  $\sigma$-model.

 Recently, two experiments, namely, NA48 at CERN and KLOE at Frascati,
have announced the possibility of measuring
the asymmetry $\Delta g_C$  and 
$\Delta g_N$ with a sensitivity of the order of $10^{-4}$,
i.e., two orders of magnitude better than at present \cite{ISTRA+}, 
 see for instance \cite{NA48} and \cite{KLOE}.
It is therefore mandatory to have 
these predictions  at  NLO in CHPT. 
 The goal of this paper is to make such predictions.

In particular, 
we have explicitly checked the one-loop results of \cite{BDP03},
we also provide the complete one-loop calculation for the electroweak octet
contribution up to ${\cal O}(e^2p^2)$  in CHPT for all the 
decays $K\rightarrow 3\pi$
and  finally, we estimate  the dominant FSI for the charged
Kaon $K\rightarrow 3\pi$ decays. We use all this to make  the first full
NLO in CHPT predictions  for the charged Kaon $K\to3\pi$ slope $g$
and decay rates CP-violating asymmetries.
We also present analytical results for all of our predictions.

Notation and definitions of the asymmetries are in Section
 \ref{notat}.
 In Section \ref{inputs} we collect the inputs that we use
for the weak counterterms in the leading and next-to-leading order
weak chiral Lagrangians.
In Section \ref{CPconserving}
 we give the CHPT predictions  at leading- and next-to-leading 
order for the decay rates and the slopes $g$, $h$ and $k$.  
We discuss the results for the CP-violating
asymmetries at leading order 
  first in Section \ref{LOresults} 
and we discuss them at NLO in Section \ref{NLOresults}.
Finally,  we give the conclusions and make comparison 
with earlier work in Section \ref{conclu}.
In Appendix \ref{LagNLO}, the $\Delta S=1$ CHPT Lagrangian used at NLO
can be found. In  Appendix \ref{Amplitudes} we give the notation
we use for  the $K \to 3 \pi$ amplitudes and the new
results at order $e^2 p^2$. In Appendix \ref{Adeltag}
 we give the analytic formulas needed for the slope $g$ and the asymmetries
$\Delta g$ at LO and NLO 
and in Appendix \ref{ANLO}  the relevant quantities  to calculate
the decay rates $\Gamma$ and  
the CP-violating asymmetries in the decay rates  $\Delta \Gamma$
 also  at LO and NLO. 
In Appendix \ref{FSI6} we give  the  analytical
results for the dominant --two-bubble--  FSI contribution
to the decays of charged Kaons and to the CP-violating 
asymmetries at NLO order, i.e. order $p^6$.

\section{Notation and Definitions}
\label{notat}

The lowest order SU(3) $\times$ SU(3)
 chiral Lagrangian describing $|\Delta S|=1$ transitions is
\ba
\label{deltaS1}
{\cal L}^{(2)}_{|\Delta S|=1}&=&
C\,F_0^6 \, e^2 \, G_E \, \tr \left( \Delta_{32} u^\dagger Q u\right)
+ C F_0^4 \left[ G_8 \, \tr \left( \Delta_{32} u_\mu u^\mu \right)
+ G_8' \tr \left( \Delta_{32} \chi_+ \right) \right.
\nonumber \\  &+& \left. 
G_{27} \, t^{ij,kl} \, \tr \left( \Delta_{ij} u_\mu \right) \,
\tr \left(\Delta_{kl} u^\mu\right) \right] + {\rm h.c.}
\ea
with
\be \label{Cdefinicion}
C= -\frac{3}{5} \frac{G_F}{\sqrt 2} V_{ud} {V^*_{us}} 
\simeq -1.07 \times  10^{-6} \, {\rm GeV}^{-2}\,.
\ee
The correspondence with the couplings $c_2$ and $c_3$ of 
\cite{KMW91,KMW90} is 
\ba
c_2=C F_0^4 \, G_8; \nonumber \\
 c_3= -\frac{\dis 1}{\dis 6} C F_0^4 \, G_{27} \, . 
\ea
$F_0$ is the chiral limit value of the pion decay
constant $f_\pi= (92.4 \pm 0.4)$ MeV, 
\ba
u_\mu \equiv i u^\dagger (D_\mu U) u^\dagger = u_\mu^\dagger \; , 
\nonumber \\
\Delta_{ij}= u \lambda_{ij} u^\dagger\; (\lambda_{ij})_{ab}\equiv
\delta_{ia} \delta_{jb}\; , \nonumber \\ 
\chi_{+(-)}= u^\dagger \chi u^\dagger +(-)u \chi^\dagger u  
\ea 
$\chi= \mbox{diag}(m_u,m_d,m_s)$ 
a 3 $\times$ 3 matrix collecting the light quark masses, 
$U\equiv u^2=\exp{(i\sqrt 2 \Phi /F_0)}$ is the exponential
representation incorporating the octet of light pseudo-scalar mesons
in the SU(3) matrix $\Phi$; 

\ba
\Phi\equiv \left(  
\begin{array}{ccc}
\frac{\dis \pi^0}{\dis \sqrt{2}} + 
\frac{\dis \eta_8}{\dis \sqrt{6}} & \pi^+ & K^+ 
\nonumber \\ 
\pi^- & -\frac{\dis \pi^0}{\dis \sqrt{2}} 
+ \frac{\dis \eta_8}{\dis \sqrt{6}} & K^0 
\nonumber \\ 
K^- & \bar{K^0} &- 2 \frac{\dis \eta_8}{\dis \sqrt{6}}  
\end{array}
\right) \, .
\ea

The non-zero components of the 
SU(3) $\times$ SU(3) tensor $t^{ij,kl}$  are
\ba
t^{21,13}=t^{13,21}=\frac{1}{3} \, ; \, &  t^{22,23}=t^{23,22}=
-\frac{\dis 1}{\dis 6} \, ;
\nonumber \\
t^{23,33}=t^{33,23}=-\frac{1}{6} \, ; \, 
&  t^{23,11}=t^{11,23}=\frac{\dis 1}{\dis 3} \, ; 
\ea
and $Q=\mbox{diag}(2/3,-1/3,-1/3)$ is a 3 $\times$ 3
matrix which collects the electric charge of the three light 
quark flavors.

We calculate  the amplitudes 
\ba
\label{defdecays}
K_2(k)&\to&\pi^0(p_1)\pi^0(p_2)\pi^0(p_3)\,,\quad [A^2_{000}]\,,\nonumber\\
K_2(k)&\to&\pi^+(p_1)\pi^-(p_2)\pi^0(p_3)\,,\quad [A^2_{+-0}]\,,\nonumber\\
K_1(k)&\to&\pi^+(p_1)\pi^-(p_2)\pi^0(p_3)\,,\quad [A^1_{+-0}]\,,\nonumber\\
K^+(k)&\to&\pi^0(p_1)\pi^0(p_2)\pi^+(p_3)\,,\quad [A_{00+}]\,,\nonumber\\
K^+(k)&\to&\pi^+(p_1)\pi^+(p_2)\pi^-(p_3)\,,\quad [A_{++-}]\,,
\ea
as well as their CP-conjugated decays
at NLO (i.e. order $p^4$ in this case) 
in  the  chiral expansion and in the isospin symmetry limit
$m_u=m_d$. We have also calculated the  contribution of the
$\order(e^2 p^2)$ electroweak octet counterterms. 
In (\ref{defdecays}) we 
 have indicated the four momentum carried by each particle
and the symbol we will use for the amplitude.
The states $K_1$ and $K_2$ are defined as
\be
K_{1(2)} \,=\,\frac{K^0-(+)\overline{K^0}}{\sqrt{2}} \, .
\ee

For the explicit form of the
Lagrangian we have used, see Appendix \ref{LagNLO}. 
Our results for the  octet and 27-plet terms fully agree with
the results found in \cite{BDP03} so that we do not write them again.
 The electroweak 
 contributions to $K \to 3 \pi$ decays  of order $e^2 p^0$ and
$e^2 p^2$ can be found 
in  Subsection \ref{ZiK3pi}  in Appendix \ref{Amplitudes}.

In this paper we  discuss CP-violating asymmetries
in the decay of the charged Kaon into three pions;
namely, asymmetries  in the slope $g$ defined as
\be \label{gdefinition}
\frac{\left|A_{K^+\to 3 \pi}(s_1,s_2,s_3)\right|^2}
{\left|A_{K^+\to 3 \pi}(s_0,s_0,s_0) \right|^2}=
1+ g \, y + h \, y^2 + k \, x^2 + {\cal O}( y  x^2,y^3) \, 
\ee
 and some asymmetries in the integrated $K^+ \to  3 \pi$ decay rates. 
Above,  we used the Dalitz variables
\ba \label{Dalitzvar}
x\equiv \frac{s_1-s_2}{m_{\pi^+}^2}& \hspace*{0.5cm} {\rm and}
\hspace*{0.5cm} &
y \equiv \frac{s_3-s_0}{m_{\pi^+}^2} \, 
\ea
with 
$s_i\equiv(k-p_i)^2$, $3s_0\equiv m_K^2 + m_{\pi^{(1)}}^2+ 
m_{\pi^{(2)}}^2+m_{\pi^{(3)}}^2$. 

The CP-violating asymmetries in the slope $g$ 
are defined as 
\ba
\label{defDeltag}
\Delta g_C \equiv
\frac{g[K^+ \to \pi^+ \pi^+ \pi^-]-g[K^-\to\pi^-\pi^-\pi^+]}
{g[K^+ \to \pi^+ \pi^+ \pi^-]+g[K^-\to\pi^-\pi^-\pi^+]}
\nonumber \\ {\rm and} \hspace*{0.5cm} 
 \Delta g_N \equiv 
\frac{g[K^+ \to \pi^0 \pi^0 \pi^+]-g[K^-\to\pi^0\pi^0\pi^-]}
{g[K^+ \to \pi^0 \pi^0 \pi^+]+g[K^-\to\pi^0\pi^0\pi^-]} \, .
\ea
 A first update at LO of these asymmetries 
was already presented in \cite{GPS03M}. 

The CP-violating asymmetries in the decay rates are defined as
\ba 
\label{defDeltaGam}
\Delta \Gamma_C \equiv
\frac{\Gamma[K^+ \to \pi^+ \pi^+ \pi^-]-\Gamma[K^-\to\pi^-\pi^-\pi^+]}
{\Gamma[K^+ \to \pi^+ \pi^+ \pi^-]+\Gamma[K^-\to\pi^-\pi^-\pi^+]}
 \nonumber \\  
 {\rm and} \hspace*{0.5cm} \Delta \Gamma_N \equiv 
\frac{\Gamma[K^+ \to \pi^0 \pi^0 \pi^+]-\Gamma[K^-\to \pi^0\pi^0\pi^-]}
{\Gamma[K^+ \to \pi^0 \pi^0 \pi^+]+\Gamma[K^-\to \pi^0\pi^0\pi^-]} \, .
\ea

In particular, we also  want
 to check the statement that with appropriate cuts one can
get one order of magnitude enhancement in $\Delta \Gamma_C$
and $\Delta \Gamma_N$ asymmetries \cite{AVI81}.

\section{Numerical Inputs for the Weak Chiral Counterterms} 
\label{inputs}

Here we collect the values of the weak  chiral counterterms that
we use in this work.

\subsection{Counterterms of the LO Weak Chiral Lagrangian} 
\label{discusscouplins}

In \cite{BDP03},  a fit to all available $K \to \pi \pi$ 
amplitudes at NLO in CHPT \cite{BPP98}  and
$K \to 3 \pi$ amplitudes  and slopes in the $K\to 3 \pi$ 
amplitudes at NLO in CHPT was done.
The result found there for the ratio of the isospin definite [0 and 2]
$K\to \pi \pi$ amplitudes to all orders in CHPT was
\be
\frac{A_0[K\to \pi\pi]}{A_2[K\to\pi\pi]}= 21.8 \, ; 
\ee
giving the infamous $\Delta I=1/2$ rule for Kaons and 
\be
\left[\frac{A_0[K\to \pi\pi]}{A_2[K\to\pi\pi]}\right]^{(2)}= 17.8 \, , 
\ee
to lowest CHPT order $p^2$. I.e., Final State Interactions  and the 
rest of higher order corrections are responsible for 22\% of the 
$\Delta I= 1/2$ enhancement 
rule. Yet most of this enhancement appears at lowest 
CHPT order! The last result is equivalent [using $F_0=87.7 $ MeV] to
\ba
\label{experiment}
\re G_8 = 6.8 \pm 0.6 & \hspace*{0.5cm} {\rm and} 
\hspace*{0.5cm} & G_{27} = 0.48 \pm 0.06\, .
\ea
In this normalization, $\re G_8=G_{27}=1$ at large $N_c$.
No information can be obtained for $ \re (e^2 G_E)$ due to its tiny 
contribution to CP-conserving amplitudes.

CP-conserving observables are fixed by 
physical meson masses, the pion decay coupling in the
chiral limit $F_0$
and the real part of the counterterms.
 To predict CP-violating asymmetries 
we  also need the values of the imaginary part of 
these couplings. Let us see what we know about them. 
At large $N_c$, all  the contributions 
to $\im G_8$ and  $\im (e^2 G_E)$  are 
factorizable and the  scheme dependences are not  under control.
The unfactorizable topologies are not included at this order and they
bring in unrelated dynamics with its new scale and scheme
dependence, so that one cannot give an uncertainty
to the large $N_c$ result for $\im G_8$ and $\im (e^2 G_E)$.
 We get
\ba \label{largeNccouplings}
\im G_8\Big|_{N_c} \!\!\!&=&\!\!\! \, 1.9   
 \, \im \tau \, , \nonumber \\
\im(e^2 G_E)\Big|_{N_c} \!\!\!&=&\!\!\!\,   -2.9    \, \im \tau \, .
\ea
In the Standard Model \cite{SCH03}
\be
\im \tau \equiv -\im 
\left( \frac{V_{td}V^*_{ts}}{V_{ud}V^*_{us}} \right) 
\simeq -(6.05\pm0.50) \times 10^{-4} \, ,  
\ee
and we used \cite{BPR95}
\be
\langle 0 | \overline q q | 0 \rangle_{\overline{\rm MS}}(2 {\rm GeV})
=-(0.018 \pm 0.004) \, {\rm GeV}^3 
\ee
which agrees with the most recent sum rule determinations
of this condensate and of light quark masses 
--see \cite{JOP02} for instance-- and the lattice light
quark masses world average \cite{WIT02}.

There have been recently advances on going beyond the leading order 
in $1/N_c$ in both couplings, $\im G_8$ and $\im (e^2 G_E)$.

In \cite{CDGM03,NAR01,BGP01}, there are recent model independent
calculations of $\im (e^2 G_E)$. The results there are valid to all
orders in $1/N_c$ and NLO in $\alpha_S$. They are obtained using 
the hadronic tau data collected by ALEPH \cite{ALEPH} and OPAL 
\cite{OPAL} at LEP. The agreement is quite good between them and their 
results can be summarized in
\be
\label{EMpenguin}
\im (e^2 G_E) = - (4.0 \pm 0.9) \, \im \tau \, , 
\ee
where the central value is an average and the error
is  the smallest one. In \cite{KPR01} it was used a Minimal
Hadronic Approximation to large $N_c$ to calculate $\im (e^2 G_E)$, 
they got
\be
\im (e^2 G_E) = -(6.7 \pm 2.0) \, \im \tau \, ,
\ee
which is also in agreement though somewhat larger. There are also lattice
results for $\im (e^2 G_E)$ both  
using domain-wall fermions \cite{domainwall} and
 Wilson fermions \cite{wilson}. All of them made the chiral limit 
extrapolations, their results are in agreement between themselves and
their average gives
\be
\im (e^2 G_E) = - (3.2 \pm 0.3) \, \im \tau \, .
\ee

There are  also results on  $\im G_8$ at NLO in $1/N_c$.
 In \cite{BP00}, the authors made a calculation
using a hadronic model which reproduced the $\Delta I=1/2$ rule for Kaons
through a very large $Q_2$ penguin-like contribution
--see \cite{BP99} for details. 
The results obtained there are
\ba
\re G_8 = 6.0 \pm 1.7 , \,\,\,   & {\rm and}& G_{27} = 0.35 \pm 0.15 \, ,
\ea
in very good agreement with the experimental results in (\ref{experiment}).

The result found in \cite{BP00} is
\be
\label{gluonpenguin}
\im G_8 = (4.4 \pm 2.2) \, \im \tau  
\ee
at NLO in $1/N_c$.
The hadronic model used there had however some drawbacks
\cite{PPR98} which have been eliminated in the
ladder resummation hadronic model in \cite{BGLP03}.
The work in \cite{BP00,BP99} will be eventually updated
using this hadronic model.

 In \cite{BP00} there was also  a determination
of $\re (e^2 G_E)$ though very uncertain.
However, since the contribution of $\re (e^2 G_E)$ is very small
in all the quantities we calculate, we take the value from 
\cite{BP00} with 100\% uncertainty and
add its contribution  to the error of those quantities.

Very recently, using a Minimal Hadronic Approximation to large $N_c$, 
the authors of \cite{HPR03} found qualitatively similar results
to those in \cite{BP00}.
I.e. enhancement toward the explanation of the $\Delta I=1/2$ rule
through $Q_2$ penguin-like diagrams and a matrix element of the gluonic
penguin $Q_6$ around three times the factorisable contribution.
The same type of enhancement though less moderate was already found
in \cite{HKPS00}.

\subsection{Counterterms of the NLO Weak Chiral Lagrangian}

To describe $K\to 3 \pi$ at NLO, 
in addition to $\re G_8$, $G_{27}$, $\re (e^2 G_E)$, $\im G_8$ and
$\im (e^2 G_E)$, we also need several other ingredients.
Namely, for the real part we need the chiral logs and the counterterms. 
The relevant counterterm combinations 
were called $\widetilde K_i$ in \cite{BDP03}. 
The chiral logs are fully 
analytically known \cite{BDP03} --we have confirmed them 
in the present work. The real part of the counterterms,
 $\re \widetilde K_i$, can be obtained from 
the fit  of the $K \to 3 \pi$ 
CP-conserving decays  to data done in \cite{BDP03}. The relation of the 
$\widetilde K_i$ counterterms and those defined in Appendix \ref{LagNLO}, 
and the values used for them are in Table \ref{tabKdef} and Table 
\ref{tabKvalues} respectively.
\TABLE{
\label{tabKdef}
\begin{tabular}{|c|c|}\hline
$\widetilde K_1$ & $ \re (G_8) (N_5^r-2N_7^r+2N_8^r+N_9^r)+G_{27}
\left(-\frac{1}{2}D_6^r\right)$\\\hline
$\widetilde K_2$ & $ \re (G_8) (N_1^r+N_2^r)+G_{27}
\left(\frac{1}{3}D_{26}^r-\frac{4}{3}D_{28}^r\right)$\\\hline
$\widetilde K_3$ & $\re (G_8) (N_3^r)+G_{27}
\left(\frac{2}{3}D_{27}^r+\frac{2}{3}D_{28}^r\right)$\\\hline
$\widetilde K_4$ & $G_{27}\left(D_4^r-D_5^r+4D_7^r\right)$\\\hline
$\widetilde K_5$ & $G_{27}\left(D_{30}^r+D_{31}^r+2D_{28}^r\right)$\\\hline
$\widetilde K_6$ & $G_{27}\left(8D_{28}^r-D_{29}^r+D_{30}^r\right)$\\\hline
$\widetilde K_7$ & $G_{27}\left(-4D_{28}^r+D_{29}^r\right)$\\\hline
$\widetilde K_8$ & $ \re(G_8)
(2N_5^r+4N_7^r+N_8^r-2N_{10}^r-4N_{11}^r-2N_{12}^r)
+G_{27}\left(-\frac{2}{3}D_1^r+\frac{2}{3}D_6^r\right)$\\\hline
$\widetilde K_9$ & $\re(G_8)
(N_5^r+N_8^r+N_9^r)+G_{27}
\left(-\frac{1}{6}D_6^r\right)$\\\hline
$\widetilde K_{10}$ & $G_{27}\left(2D_2^r-2D_4^r-D_7^r\right)$\\\hline
$\widetilde K_{11}$ & $G_{27} D_7^r $\\\hline
\end{tabular}
\caption{Relevant combinations of the octet $N_i^r$ 
and 27-plet $D_i^r$ weak counterterms for $K\to 3\pi$ 
decays.}     }
\TABLE{
\label{tabKvalues}
\begin{tabular}{||c|c|c||}\hline
 & $\re \widetilde K_i(M_\rho)$ from \cite{BDP03}& 
$\im \widetilde K_i(M_\rho)$  from (\ref{assum1})\\  
\hline\hline
$\widetilde K_2(M_\rho)$ & $0.35\pm0.02$
& $\lbrack 0.31 \pm 0.11 \rbrack\, \im \tau$ \\ 
\hline
$\widetilde K_3(M_\rho)$ & $0.03\pm0.01$ 
& $\lbrack 0.023\pm 0.011\rbrack \,\im \tau$ \\
\hline
$\widetilde K_5(M_\rho)$ & $-(0.02\pm0.01)$& 
$0$\\ \hline
$\widetilde K_6(M_\rho)$ & $-(0.08\pm0.05)$ & $0$ \\\hline
$\widetilde K_7(M_\rho)$ & $0.06\pm0.02$ & $0$ \\\hline
\end{tabular}
\caption{Numerical inputs used for the weak counterterms of 
order $p^4$. 
The values of $\re \widetilde K_i$ and $\im \widetilde K_i$ 
which do not appear are zero. For explanations, see the text.} }

For the imaginary parts
at NLO, we need $\im G_8'$ in addition to $\im G_8$ and $\im (e^2 G_E)$.
To the best of our knowledge, there is just one calculation  at NLO 
in $1/N_c$ at present \cite{BP00}. The results found there,
using the same hadronic model discussed above, are
\ba
\label{G8pri}
\re G_8' = 0.9 \pm 0.1  \,\, \,  
  & {\rm and} & \im G_8' = (1.0 \pm 0.4) \,
\im \tau \, .
\ea
The imaginary part of the order $p^4$ counterterms, $\im \widetilde K_i$,
is much more problematic.  They cannot be obtained from data and 
there is no available NLO in $1/N_c$ calculation for them.

One can use several approaches to get the order of magnitude
and/or the signs of $\im \widetilde K_i$.  Among these approaches  are
factorization plus meson dominance \cite{EGPR89}.
 If one uses factorization, 
one needs couplings of order $p^6$ from the strong chiral Lagrangian
for some of the $\widetilde K_i$ counterterms, see also \cite{PPI01}.
Not very much is known about these $\order(p^6)$ couplings though. 
 One can use  Meson Dominance 
to  saturate them  but it is not clear that this procedure
 will be in general a good 
estimate.  See for instance \cite{KN01}  for some detailed analysis of 
some order $p^6$  strong counterterms 
 obtained  at large $N_c$ using also short-distance QCD constraints 
and comparison with meson  exchange saturation. 
See also \cite{CENP03}  for a very recent estimate of some relevant
order $p^6$ counterterms in the strong sector using 
Meson Dominance and factorization.

Another more ambitious procedure to predict 
the necessary NLO weak counterterms is to combine short-distance QCD, 
large $N_c$ constraints plus other chiral constraints
and some phenomenological inputs
to construct the relevant $\Delta S=1$ Green functions, see
\cite{BGLP03,KN01,KPR00}. This last program has not yet been
used systematically to get all the $\Delta S=1$
counterterms at NLO.

We will follow here more naive approaches
that will be enough for our purpose of estimating the effect of the 
unknown counterterms.
 We can assume   that the ratio of the real to the imaginary parts
is  dominated by the same strong dynamics at LO and NLO 
in CHPT, therefore
\be
\label{assum1}
\frac{\im \widetilde K_i}{\re \widetilde K_i}
\simeq \frac{\im G_8}{\re G_8} 
\simeq \frac{\im G_8'}{ \re G_8'} \simeq (0.9 \pm 0.3) \, \im \tau  \, ,
\ee
if we use (\ref{gluonpenguin}) and (\ref{G8pri}).
The results obtained under these  assumptions for the imaginary 
part of the $\widetilde K_i$ counterterms are written in 
Table \ref{tabKvalues}. 
In particular, we set to zero those $\im \widetilde K_i$
whose corresponding $\re \widetilde K_i$ are 
set also to zero in the fit to CP-conserving amplitudes
done in \cite{BDP03}.
Of course, the relation above can only be applied to those 
$\widetilde K_i$ couplings
with non-vanishing imaginary part.
Octet dominance to order $p^4$ is a further
assumption implicit in (\ref{assum1}).
The second equality  in (\ref{assum1})
 is well satisfied by the model calculation
in (\ref{G8pri}). 

The values of $\im \widetilde K_i$
obtained using (\ref{assum1}) 
 will allow us to check the counterterm dependence
of the CP-violating asymmetries. They will also provide us  a good 
estimate of the counterterm contribution to the CP-violating asymmetries
that we are studying. 

We can get a second  piece of information   from the 
variation of the amplitudes when $\im \widetilde K_i$ are put to zero 
and the remaining scale dependence is varied between 
$M_\rho$ and 1.5 GeV.
We use in this case the known scale dependence of 
$ \re \widetilde K_i$ together with their 
absolute value  at the scale $\nu=M_\rho$ from \cite{BDP03}.

\section{CP-Conserving Observables} 
\label{CPconserving}

Here we give the results for the CP-conserving 
slopes $g_C$, $h_C$, and $k_C$
  and the decay rate $\Gamma_C$ of $K^+ \to \pi^+ \pi^+ \pi^-$
and slopes $g_N$, $h_N$, and $k_N$
  and decay rate $\Gamma_N$ of $K^+ \to \pi^+ \pi^0 \pi^0$
within CHPT at LO and NLO.
These results are not new --see \cite{BDP03} and references therein--
but we want to give them again, first as  a check of our analytical 
results and second, to recall the  kind of corrections that one 
expects in the CP-conserving quantities from LO to NLO
for the different observables.

We will use the values of $\re G_8$ and $G_{27}$ in (\ref{experiment}),
and disregard the EM corrections since we
are in the isospin limit  and  they are much smaller than the
octet and  27-plet contributions. For the real part of
the NLO counterterms, we will use the results from a fit to data 
in \cite{BDP03}. So, really these are just checks.

The  values of the NLO counterterms given in \cite{BDP03} were fitted 
without including CP-violating 
contributions in the amplitudes, i.e., taking 
the coupling $G_8$ and the counterterms themselves as real quantities.
The inclusion of an imaginary part for these couplings
does not affect significantly the CP  conserving observables.

To be consistent with the fitted values 
of the counterterms  of the $\order(p^4)$
Lagrangian   we do not consider any 
$\order (p^6)$ contribution to the amplitudes in this section.  
Indeed,  these  counterterms,
 fixed with the use of experimental data and order $p^4$ formulas, 
do contain the  effects of higher order contributions.
We also use the same conventions used in \cite{BDP03}
 for the pion masses, i.e., we use  the average final state
pion mass which for $K^+\to \pi^+\pi^+\pi^-$ is
$m_\pi=$ 139 MeV and for $K^+\to \pi^0\pi^0\pi^+$ is
$m_\pi=$ 137 MeV.
In the following subsections we provide analytic formulas
 at LO  and in  Tables \ref{tabCPcons}  and \ref{tabslopes}
we give the numerical results.

\subsection{Slope  $g$} \label{slopeg}

The slope $g$ is defined in equation (\ref{gdefinition}). 
We give here  the results for
\ba
 \, g_C &\equiv&\frac{1}{2} 
\Big\{g[K^+ \to \pi^+ \pi^+ \pi^-] + g[K^- \to \pi^- \pi^- \pi^+]
\Big\}
\nonumber\\
 \, {\rm and} \, \, \, g_N &\equiv& \frac{1}{2}
\Big\{g[K^+ \to \pi^0 \pi^0 \pi^+]+g[K^-\to \pi^0\pi^0\pi^-] \Big\} .
\ea
Without including  the tiny CP-violating effects  
$g[K^+]_{\rm LO}=g[K^-]_{\rm LO}$, 
\ba \label{gLOCN}
g_{C}^{\rm LO} &=&  \, \, -3\mpd
\frac{3\re G_8-13G_{27}}{\mkd\left(3\re G_8 + 2G_{27}\right)+9F_0^2
\re \left(e^2G_E\right)}\, 
,\nonumber\\
g_{N}^{\rm LO} \, \, &=& \, \, 3\frac{\mpd}{(\mkd-\mpd)}
\frac{(19\mkd-4\mpd)G_{27}+6(\mkd-\mpd) \re G_8+9F_0^2
\re \left(e^2G_E\right)}
{\mkd\left(3\re G_8 + 2G_{27}\right)+9F_0^2\re \left(e^2G_E\right)} .
\nonumber\\
\ea
The value for $\re (e^2 G_E)$ is not very well known. 
However 
its contribution turns out to be negligible and for numerical purposes
we take the result for 
$\re (e^2 G_E)$  from \cite{BP00} with 100\% uncertainty.
We do not consider its contribution for the central values 
in Table \ref{tabCPcons} and  we add its effect to the quoted 
error.  In addition, the quoted uncertainty for $g^{\rm LO}_{C}$
 and $g^{\rm LO}_{N}$ 
contains  the uncertainties from
$\re G_8$ and $G_{27}$ in (\ref{experiment}).

The analytical NLO formulas are in (\ref{AgNLO}).
It is interesting to observe the impact of the counterterms 
so that we calculate also the slopes at NLO with $\widetilde K_i=0$,
see Table \ref{tabCPcons}. The contribution of the counterterms
at $\mu=M_\rho$ is relatively small for $g_C$ and $g_N$, see Table 
\ref{tabCPcons}.
\TABLE{
\label{tabCPcons}
\begin{tabular}{||c|c|c|c|c||}\hline
 &$ g_C$ & $\Gamma_C \; (10^{-18}\ {\rm GeV})$&
$ g_N$ & $\Gamma_N \; (10^{-18}\ {\rm GeV})$\\  \hline\hline
LO & $-0.16\pm0.02$ &$1.2\pm0.2$ &$0.55\pm 0.04$  
& $0.37\pm 0.07$ \\ \hline
NLO,$\widetilde K_i(M_\rho)$ &&&&\\
 from Table \ref{tabKvalues}&
$-0.22 \pm 0.02$ & $3.1 \pm 0.6$ & $0.61 \pm 0.05$ & $0.95 \pm 0.20$ 
\\\hline
NLO,&&&& \\
 $\widetilde K_i(M_\rho)=0$
 & $-0.28 \pm 0.03$ &$1.3 \pm 0.4$ & $0.80 \pm 0.05$ &  $0.41 \pm 0.12$ 
\\\hline
PDG02 &
$-0.2154\pm 0.0035$ & $2.97\pm 0.02$ & $0.652\pm 0.031$ 
& $0.92\pm0.02$\\\hline
 ISTRA+ & -- & -- & $0.627\pm 0.011$ 
& --\\\hline
 KLOE & -- & -- & $0.585\pm0.016$
& $0.95\pm0.01$\\\hline
\end{tabular}
\caption{CP conserving predictions for the slope $g$ and the decay rates. 
The theoretical errors come from the 
variation in the inputs parameters discussed in Section \ref{inputs}.
In the last three lines, we give the experimental 2002 world
average from PDG \cite{PDG02}, and the recent results
from ISTRA+ \cite{ISTRA+} and the preliminary ones from KLOE \cite{KLOE} 
which are not included in \cite{PDG02}.}}

\subsection{Slopes $h$ and $k$}

We can also predict  the slopes $h_{C(N)}$ and $k_{C(N)}$ 
defined  in (\ref{gdefinition}).
At LO,  the slope $k_{C}$  for $K^+ \to \pi^+ \pi^+ \pi^-$
and the slope $k_N$ for $K^+ \to \pi^0 \pi^0 \pi^+$  are identically zero
and the corresponding slopes $h_{C(N)}$  are equal to $g_{C(N)}^2/4$. 
The NLO results are written in Table \ref{tabslopes} 
together with the 
slopes obtained when the counterterms $\widetilde K_i$ are switched off
at $\mu=M_\rho$. We can see 
that  the slopes $h_{C(N)}$ and $k_{C(N)}$
 are dominated by the counterterm 
contribution contrary to what happened with $g_{C(N)}$ which get
the main contributions  at LO. 
\TABLE{
\label{tabslopes}
\begin{tabular}{||c|c|c|c|c||}\hline
 &$ h_C$ & $k_C $&
$ h_N$ & $k_N$\\  \hline\hline
LO & $0.006 \pm 0.001$ &$0$ &$0.075 \pm 0.003$  
& $0$ \\ \hline
NLO,$\widetilde K_i(M_\rho)$ &&&&\\
from Table \ref{tabKvalues}&
$0.012 \pm 0.005$ & $-0.0054 \pm 0.0015$ & $0.069 \pm 0.018$ 
& $0.004 \pm  0.002$ \\\hline
NLO, &&&&\\
$\widetilde K_i(M_\rho)=0$
 & $0.04 \pm 0.01$ & $0.0004 \pm 0.0025$ & $0.15 \pm 0.05$
&$0.008 \pm 0.002$\\\hline
PDG02 &$0.012\pm0.008$ & $-0.0101\pm0.0034$ & $0.057\pm 0.018$ 
& $0.0197\pm0.0054$\\\hline
 ISTRA+ & -- & -- & $0.046\pm 0.013$ 
& $0.001\pm 0.002$ \\\hline
 KLOE & -- & -- & $0.030\pm0.016$ 
& $0.0064\pm0.0032$ \\\hline
\end{tabular}
\caption{CP conserving predictions for the slopes $h$ and $k$. 
The theoretical errors come from the
variation in the inputs parameters discussed in Section \ref{inputs}.
In the last three lines, we give the experimental 2002 world
average from PDG \cite{PDG02}, and the recent results
from ISTRA+ \cite{ISTRA+}  and the preliminary ones from
KLOE \cite{KLOE}
which are not included in \cite{PDG02}.}}

\subsection{Decay Rates}

The decay rates $K \to 3 \pi$ with two identical
pions can be written as 
\be \label{Gammadef}
\Gamma_{ijl} \,\equiv \frac{1}{512 \pi^3 m_K^3}\,
\int_{s_{3min}}^{s_{3max}} ds_3\int_{s_{1min}}^{s_{1max}} ds_1\,
|A(K\rightarrow \pi^i\pi^j\pi^l)|^2 ,
\ee
with 
\ba
\label{eq:extrs} 
s_{1max}&=&(E_j^*+E_l^*)^2-\left(\sqrt{E_j^{*2}-m_j^2}
-\sqrt{E_l^{*2}-m_l^2}\right)^2\, ,\nonumber\\
s_{1min}&=&(E_j^*+E_l^*)^2-\left(\sqrt{E_j^{*2}-m_j^2}
+\sqrt{E_l^{*2}-m_l^2}\right)^2 \, ,\nonumber\\
s_{3max}&=&(m_K-m_l)^2\quad{\rm and}\quad s_{3min}=(m_i+m_j)^2\,.
\ea
The energies $E_j^*=(s_3-m_i^2+m_j^2)/(2\sqrt{s_3})$ and 
$E_l^*=(m_K^2-s_3-m_l^2)/(2\sqrt{s_3})$ are those of the pions $\pi^j$ and 
$\pi^l$ in the $s_3$ rest frame.
It is useful  to define 
\ba
\label{defACN}
|A_C|^2&=&\frac{1}{2}\Big\{
|A\left(K^+\rightarrow \pi^+\pi^+\pi^-\right)|^2
+ |A\left(K^-\rightarrow \pi^-\pi^-\pi^+\right)|^2\Big\}\,,\nonumber\\
|A_N|^2&=&
\frac{1}{2} \Big\{|A\left(K^+\rightarrow \pi^0\pi^0\pi^+\right)|^2
+ |A\left(K^-\rightarrow \pi^0\pi^0\pi^-\right)|^2 \Big\}\,.
\ea

At LO and again disregarding the tiny CP-violating effects we get
\ba
\label{GammaLO}
&&\,|A_C^{\rm LO}|^2\,\equiv \, 
|A_{++-}^{\rm LO}|^2\,=\,|A_{--+}^{\rm LO}|^2\,=\, \nonumber \\ 
&& |C|^2 \times \left| \re G_8 \left(s_3-\mkd-\mpd\right) \,+\,
\frac{G_{27}}{3}\left(13\mpd+3\mkd-13s_3\right)\,
+\,\re\left(e^2G_E\right)(-2F_0^2)\right|^2\, , \nonumber\\
&&\,|A_{N}^{\rm LO}|^2\,\equiv\,
|A_{00+}^{\rm LO}|^2\,=\,|A_{00-}^{\rm LO}|^2\,= \, |C|^2 \,
\Big| \re G_8 \left(\mpd-s_3\right) \nonumber\\
&&\hspace{2 cm}+
\frac{G_{27}}{6(\mkd-\mpd)}\left( 5m_K^4+19\mpd\mkd
-4m_{\pi}^4+s_3(4\mpd-19\mkd)\right)\nonumber\\ 
&&  \hspace{2 cm}+\re\left(e^2G_E\right)
\frac{F_0^2}{2(\mkd-\mpd)}\left(5\mpd-\mkd-3s_3\right)\Big|^2\,.
\nonumber \\
\ea
 The amplitudes $|A_{C(N)}|^2$ needed for the NLO prediction are in  
(\ref{eqANLO}) in Appendix \ref{ANLO}.

The  results for $\Gamma_C$ and $\Gamma_N$ at LO and NLO
 are in Table \ref{tabCPcons}.
 The contribution of $\re (e^2 G_E)$ is  very small 
(around 1\%) and we  include  it  in the final uncertainty  as in 
Section \ref{slopeg} together with the rest of input uncertainties.
We have also  included  in Table \ref{tabCPcons}
the results  with  the counterterms $\widetilde K_i=0$ at $\mu=M_\rho$.
 We can conclude from them that the decay widths are 
strongly dependent  on  the NLO counterterms contribution.

\section{CP-Violating Predictions at Leading Order} 
\label{LOresults}

 The numerators of the asymmetries in (\ref{defDeltag}) and 
(\ref{defDeltaGam}) are proportional to strong phases times the
real part of the squared amplitudes. At LO in CHPT, the strong phases 
start at one-loop and  are order $p^4/p^2$ while the real part of the 
squared  amplitudes 
are order $(p^2)^2$. The denominators are proportional 
to the real part of the squared amplitudes 
which are order $(p^2)^2$,  so the asymmetries (\ref{defDeltag}) 
and (\ref{defDeltaGam}) for
 the slope $g$ and decay rates $\Gamma$ are order $p^2$ in CHPT.

We have checked that
the effect of   $\re (e^2 G_E)$  is very small
also for the $\Delta g$ and $\Delta \Gamma$ asymmetries. 
 For the numerics, we have used
$\re (e^2 G_E)=0$  and used the value in \cite{BP00} with 
100\%  variation to estimate its contribution
which we have added to the 
quoted final uncertainty of the asymmetries.
For the  $\re G_8$ and $G_{27}$ we have used always 
the values in (\ref{experiment}). 
For $\im G_8$ and $\im (e^2 G_E)$, we have used
two sets of inputs; namely,  the large $N_c$ limit
predictions in (\ref{largeNccouplings})
and the values  in (\ref{gluonpenguin}) and
(\ref{EMpenguin}). 
For the pion masses we have used the same convention
used in \cite{BDP03} and given here in Section \ref{CPconserving}.
The results are reported  in Table \ref{tabLO}.

\subsection{CP Violating Asymmetries in the Slope $g$} \label{LODeltag}

At LO, the CP-violating asymmetries in the slope $\Delta g_{C(N)}$
can be written as \cite{GPS03M} 
\be \label{DeltagLOnum}
\Delta g_{C(N)}^{\rm LO}
 \simeq \frac{m_K^2}{F_0^2} \, B_{C(N)} \, \im G_8 +
D_{C(N)} \, \im (e^2 G_E) \, , 
\ee
where the functions $B_{C(N)}$ and $D_{C(N)}$ only depend
on $\re G_8$, $G_{27}$, $m_K$ and $m_{\pi}$  and can be found in 
(\ref{ACBC}) and (\ref{ANBN}) in Appendix \ref{Adeltag}. Numerically, 
\ba
\label{eq:gLOeff}
\Delta g_{C}^{\rm LO}
 \simeq \left \lbrack 1.16 \, \im G_8 -  0.12\, \im (e^2 G_E) 
\right \rbrack \times 10^{-2}\, ,
\nonumber \\ 
\Delta g_{N}^{\rm LO}
 \simeq - \left \lbrack 0.52 \, \im G_8 +  0.063\, \im (e^2 G_E) 
\right \rbrack \times 10^{-2}\, .
\ea
\TABLE{
\label{tabLO}
\begin{tabular}{||c|c|c|c|c||}\hline
&$ \Delta g_C^{\rm LO}(10^{-5})$ &$ 
\Delta \Gamma_C^{\rm LO}(10^{-6})$
&$ \Delta g_N^{\rm LO}(10^{-5})$ &$ \Delta
 \Gamma_N^{\rm LO}(10^{-6})$\\\hline\hline
(\ref{largeNccouplings}) 
& $-1.5   $ &$ -0.2   $ & $0.5   $ & $0.8   $\\
(\ref{gluonpenguin}) and (\ref{EMpenguin})& $-3.4\pm 2.1$&$-0.6\pm 0.4$&
$1.2\pm 0.8$ &$2.0\pm 1.3$\\
\hline
\end{tabular}
\caption{CP-violating predictions  at LO in the chiral
  expansion.  The details of the calculation are  
in Section \ref{LOresults}. The inputs used  for $\im G_8$ and
$\im (e^2 G_E)$ are in the first column.
The difference between $\Delta g_C^{\rm LO}$ here and the one 
reported in \cite{GPS03M} comes from updating the values
of $\re G_8$ and $G_{27}$ from \cite{BDP03}.
The error in the first line is not reported for the reasons 
explained in Section \ref{inputs}.}}
 \ From (\ref{eq:gLOeff}) and  the inputs 
discussed in Section  \ref{discusscouplins}
we conclude that the asymmetries 
$\Delta g_{C (N)}$ are poorly sensitive to $\im (e^2 G_E)$.
This fact makes  an accurate enough measurement of these asymmetries 
 very interesting to check if  
$\im G_8$ can be as large as predicted in \cite{BP00,HPR03,HKPS00}.
It also makes these CP-violating asymmetries
complementary to the direct CP-violating parameter $\varepsilon'_K$
where there is a  cancellation between the $\im G_8$ 
and $\im (e^2 G_E)$ contributions. 

\subsection{CP-Violating Asymmetries in the Decay Rates}
\label{5.2}

The observables we study here were   defined in
(\ref{defDeltaGam}). We can write them again as follows
\ba
\label{eq:dGLO}
\Delta\Gamma_{C(N)}&=&
\frac{\int_{s_{3min}}^{s_{3max}}ds_3\int_{s_{1min}}^{s_{1max}}ds_1
   \, \Delta |A_{C(N)}|^2}{
\int_{s_{3min}}^{s_{3max}}ds_3\int_{s_{1min}}^{s_{1max}}ds_1
 \, |A_{C(N)}|^2 }
\ea
 where the extremes of integration are 
in (\ref{eq:extrs}), the quantities $|A_{C(N)}|^2$ were defined
in (\ref{defACN}) and $\Delta |A_{C(N)}|^2$ are defined by
\ba \label{Deltapm}
\Delta |A_C|^2&=& \frac{1}{2}
\Big\{ |A\left(K^+\rightarrow \pi^+\pi^+\pi^-\right)|^2
- |A\left(K^-\rightarrow \pi^-\pi^-\pi^+\right)|^2\Big\} \,,\nonumber\\
\Delta |A_N|^2&=&\frac{1}{2} 
\Big\{|A\left(K^+\rightarrow \pi^0\pi^0\pi^+\right)|^2
-|A\left(K^-\rightarrow \pi^0\pi^0\pi^-\right)|^2\Big\}\,.
\ea

At LO we get,  
\ba \label{DGammaLO}
\Delta |A_{C(N)}^{\rm LO}|^2 &=& 2 \,
\Bigg \lbrack \im G_8 \left\{ \,G_{27}\left(B_{8}^{(2)}C_{27}^{(4)}
-B_{27}^{(2)}C_8^{(4)}\right)\,\right.
\nonumber \\ && \left.
\hspace{1 cm}+\,\re \left(e^2G_{E}\right)\left(B_{8}^{(2)}C_{E}^{(4)}
-B_{E}^{(2)}C_8^{(4)}\right)\right\}\nonumber\\
&&+\im \left(e^2G_E\right) \,\left\lbrace \re G_8 \left(B_E^{(2)}C_8^{(4)}
-B_8^{(2)}C_E^{(4)}
\right)\right.
\nonumber \\ && \left. 
\hspace{1 cm}+ G_{27}\left(B_E^{(2)}C_{27}^{(4)}-B_{27}^{(2)}C_E^{(4)}\right)
\right\rbrace\hspace{1.5 cm}\Bigg \rbrack\, ,
\ea
where we  do not show explicitly the  $s_j$ 
dependence of the functions $B_i^{(2)}$ and $C_i^{(4)}$
nor the subscript $C$ or $N$  in $B_i^{(2)}$ and  $B_i^{(4)}$
for the sake of simplicity.
The analytical expressions for  the functions $B_i^{(2)}$ and $C_i^{(4)}$
are reported in Appendix \ref{ANLO}.

In (\ref{DGammaLO}),  we have
used  consistently the  LO  result for  the denominator of 
(\ref{eq:dGLO}) though its value is very different
from the experimental number, see Table \ref{tabCPcons}.
 
The numerics  for the asymmetries in the decay rates are
\ba
\Delta \Gamma_{C}^{\rm LO}
 \simeq \left\lbrack 0.24  \, \im G_8 +  0.03\, 
\im (e^2 G_E) \right\rbrack \times 10^{-3} \, ,
\nonumber \\ 
\Delta \Gamma_{N}^{\rm LO} \simeq -\left\lbrack 0.88  \, \im G_8 + 
0.13\, \im(e^2 G_E)  \right\rbrack \times 10^{-3}\, .
\ea
The results using the two sets of inputs discussed in Section
\ref{inputs} for
$\im G_8$ and $\im (e^2 G_E)$ are reported in Table \ref{tabLO}.
The asymmetries  in the width are also poorly sensitive to 
$\im (e^2 G_{E})$
thus also their accurate measurement
 will provide important information on  $\im G_8$.

In \cite{AVI81},  it was noticed that the asymmetry 
$\Delta \Gamma_{C}$ increases if a cut on the energy of the 
pion with charge opposite that of the decaying Kaon is made. Afterward, 
 the authors in  \cite{GRW86}  claimed that if this cut is made 
at $s_3=1.1\times 4\mpd$, the asymmetry 
 is enhanced by one  order of magnitude.
We checked that the decay rate asymmetry $\Delta \Gamma_C$
at LO changes from its value in Table \ref{tabLO} 
 to $\Delta \Gamma_{C}=-5.6\times10^{-6}$, i.e.
one order of magnitude enhancement when we perform such a cut in the 
integration, 
in agreement with \cite{GRW86}. It remains to see if this
enhancement persists  at NLO and how feasible is 
to perform this cut at the experimental side.
We will come back to this issue in the conclusions
in Section \ref{conclu}. This enhancement
does not occur for $\Delta \Gamma_N$.

\section{CP-Violating Predictions at Next-to-Leading Order} 
\label{NLOresults}

At NLO one needs the real parts at order $p^4$, 
i.e. at one-loop, for which we have
the exact expression, see Appendix \ref{Amplitudes}. 
To make the full discussion about CP-violating asymmetries at NLO
in CHPT we also need the FSI at order $p^6$
that would imply to calculate  $K \to 3 \pi$ amplitudes
at two-loops. However, one can use the optical  theorem 
and the one-loop and tree-level  $\pi \pi$ scattering and $K \to
3 \pi$ results 
 to get the imaginary part  of the dominant two-bubble
contributions. The results for these dominant two-bubble
FSI are presented in the next subsection.

\subsection{Final State Interactions at NLO}
\label{6.1}

Though the complete analytical FSI at NLO are unknown at present, 
 one can do a very good job using the known results at order 
$p^2$ and order $p^4$ for $\pi \pi$ scattering and for $K \to 3 \pi$ 
together with the optical theorem
to get analytically the  order $p^6$ imaginary parts that come
from two-bubbles. 
These contributions are expected to be dominant to a very good accuracy.
 We are disregarding  three-body re-scattering since they cannot be 
written as a bubble resummation. One can expect them to be rather 
small being suppressed by the available phase space \cite{DIPP94}.

Making use of the Dalitz variables defined in (\ref{Dalitzvar}) the 
amplitudes in (\ref{defdecays}) 
[without isospin breaking terms] 
can be written as expansions in powers of  $x$ and $y$, 
\ba \label{amp1}
A_{++-} &=& (-2\alpha_1+\alpha_3)\,-\,(\beta_1-\frac{1}{2}\beta_3+\sqrt 3
\gamma_3)\,y \,+\,\order(y^2,x)\, ,\nonumber\\
A_{00+} &=& \frac{1}{2}(-2\alpha_1+\alpha_3)\,-\,(-\beta_1
+\frac{1}{2}\beta_3+\sqrt 3\gamma_3)\,y  \,+\,\order(y^2,x)\,,\nonumber\\
A^2_{+-0} &=& (\alpha_1+\alpha_3)^R\,- \,(\beta_1+\beta_3)^R\,y
 \,+\,\order(y^2,x)\, ,\nonumber\\
A^1_{+-0} &=&  (\alpha_1+\alpha_3)^I\,- \,(\beta_1+\beta_3)^I\,y
 \,+\,\order(y^2,x)\, ,\nonumber\\
A^2_{000} &=& 3\,(\alpha_1+\alpha_3)^R \,+\,\order(y^2,x) \, ,\nonumber\\
A^1_{000} &=& 3\,(\alpha_1+\alpha_3)^I \,+\,\order(y^2,x) \, ,
\ea
where the parameters $\alpha_i$, $\beta_i$ and $\gamma_i$ are  
functions of the 
pion and Kaon masses, $F_0$, the lowest order $\Delta S=1$ Lagrangian
couplings $G_8$, 
$G_8'$, $G_{27}$, $G_E$ and the counterterms  appearing 
at order $p^4$, i.e., $L_i's$, $\widetilde K_i's$. 
We do not add here
EM corrections since we expect them to be small and of the same size of 
isospin breaking effects in quark masses which we have not considered.
 The $\order ( e^2 p^0)$ and $\order (e^2 p^2)$
contributions can be found in Subsection \ref{ZiK3pi}.

 In (\ref{amp1}), superindices $R$  and $I$ mean that either  the 
real part of the counterterms or their imaginary part 
appear, respectively. In the remainder,  the 
superscript ${(+-0)}$ will refer to the 
amplitude $A(K^0\to\pi^+\pi^-\pi^0)= 
(A^2_{+-0}+A^1_{+-0})/\sqrt{2})$, that is proportional to 
the full couplings and not 
only to the real or the imaginary part of such couplings.

If we do not consider FSI, the complex parameters $\alpha_{i}^{{\rm NR}}$, 
$\beta_{i}^{{\rm NR}}$ and $\gamma_{i}^{{\rm NR}}$ 
--with the superscript ${\rm NR}$ meaning that 
re-scattering effects have not been included-- can be written at NLO 
in terms of the order $p^2 $ and $p^4$ 
counterterms and the constants  $B_{i,0(1)}=
B_{i,0(1)}^{(2)}+B_{i,0(1)}^{(4)}$ 
and $H^{(4)}_{i,0(1)}$  defined in (\ref{F4F6def}),  
(\ref{functionsNLO}) and (\ref{yexpansion}). They can be obtained
from Appendix \ref{ANLO} by expanding the corresponding functions
$B_i$, $C_i$ and $H_i$ as in (\ref{yexpansion}). We get
\ba \label{alphas}
\alpha_{1}^{\rm NR} &=&  \ -G_8\,\frac{1}{2}\,B_{8,0}^{(++-)}
\,+\,\frac{1}{3}\,\sum _{i=27,E}G_{i}\,\left(B_{i,0}^{(+-0)}-
B_{i,0}^{(++-)}\right)\nonumber\\ 
&&\hspace{2 cm}+  \frac{1}{3}\,\sum_{i=1,11}\left(H_{i,0}^{(4)(+-0)}-
H_{i,0}^{(4)(++-)}\right)\widetilde K_i 
  \, ,
\nonumber\\
\alpha_{3}^{\rm NR} &=& \sum _{i=27,E} G_{i}\,\frac{1}{3}
\,\left(B_{i,0}^{(++-)}+2B_{i,0}^{(+-0)}\right) 
+  \frac{1}{3}\,\sum_{i=1,11}\left(H_{i,0}^{(4)(++-)}
+2H_{i,0}^{(4)(+-0)}\right)\widetilde K_i  \, , \nonumber \\
\beta_{1}^{\rm NR} &=&   -G_8\,B_{8,1}^{(++-)}
\,-\,\frac{1}{3}\, \sum _{i=27,E} G_{i}\left(B_{i,1}^{(+-0)}+
B_{i,1}^{(++-)}-B_{i,1}^{(00+)}\right)\nonumber\\
&-& \frac{1}{3}\,\sum_{i=1,11}\left(H_{i,1}^{(4)(+-0)}
+H_{i,1}^{(4)(++-)}-H_{i,1}^{(4)(00+)}\right)\widetilde K_i \, ,\nonumber\\
\beta_{3}^{\rm NR} &=& \frac{1}{3}\,   \sum _{i=27,E} G_{i}
\left(B_{i,1}^{(++-)}-B_{i,1}^{(00+)}-2B_{i,1}^{(+-0)}\right)\nonumber\\  
&&+  \frac{1}{3}\,\sum_{i=1,11}\left(H_{i,1}^{(4)(++-)}
-H_{i,1}^{(4)(00+)}-2H_{i,1}^{(4)(+-0)}\right)\widetilde K_i\,,\nonumber\\
\sqrt 3 \gamma_{3}^{\rm NR} &=&  
- \frac{1}{2}\,\sum _{i=27,E} G_{i}\left( B_{i,1}^{(++-)}
+ B_{i,1}^{(00+)}\right)
- \frac{1}{2}\,\sum_{i=1,11}\left(H_{i,1}^{(4)(++-)}
+H_{i,1}^{(4)(00+)}\right)\widetilde K_i\,.
\ea

At LO the expressions above give
\ba 
\alpha_{1}^{\rm LO} &=& iC\left \lbrack G_8 \frac{\mkd}{3}\,+\, 
G_{27}\frac{\mkd}{27}\,+ \,e^2G_E\frac{2}{3} F_0^2\right\rbrack\,,
\nonumber\\
\alpha_{3}^{\rm LO} &=& iC\left \lbrack 
-G_{27}\frac{10\mkd}{27}\,-\,e^2G_E\frac{2}{3} F_0^2\right\rbrack\,,
\nonumber\\
\beta_{1}^{\rm LO} &=& iC\left \lbrack -G_8 \mpd\,-\, 
G_{27}\frac{\mpd}{9}\right\rbrack\,,
\nonumber\\
\beta_{3}^{\rm LO} &=& iC\left \lbrack -G_{27} \frac{5\mpd}{18(\mkd-\mpd)}
\left(5\mkd-14\mpd\right)\,
+ \,e^2G_E F_0^2\frac{3\mpd}{2(\mkd-\mpd)}\right\rbrack\,,\nonumber\\
\sqrt{3}\gamma_{3}^{\rm LO} 
&=& iC\left \lbrack G_{27} \frac{5\mpd}{4(\mkd-\mpd)}
\left(3\mkd-2\mpd\right)\,
+ \,e^2G_E F_0^2\frac{3\mpd}{4(\mkd-\mpd)}\right\rbrack\, ,
\ea
with the constant $C$ defined in (\ref{Cdefinicion}).

The strong FSI  mix the two final states with isospin $I=1$
and leaves unmixed  the isospin 
$I=2$ state. The mixing in the isospin $I=1$ 
decay amplitudes is taken into account by introducing the
strong re-scattering 2 $\times$ 2 matrix $\mathbb{R}$ \cite{DIP91}. 
The amplitudes in (\ref{defdecays}) including  the FSI effects 
can be written as follows at all orders,  
\ba
\label{Rescat}
T_c\left(\begin{array}{c}A_{++-}^{(1)}
\\A_{00+}^{(1)}\end{array}\right)_{\rm R} &=& 
\Big(\, {\mathbb{I}} 
+i\,\mathbb{R}\,\Big) T_c\left(\begin{array}{c}A_{++-}^{(1)}
\\A_{00+}^{(1)} \end{array}\right)_{\rm NR} \, ,\nonumber\\
T_n\left(\begin{array}{c}A_{+-0}^{(2)}
\\ A_{000}^{(2)} \end{array}\right)_{\rm R} &=& 
\Big(\,{\mathbb{I}}+i\,\mathbb{R}
\,\Big)T_n\left(\begin{array}{c}A_{+-0}^{(2)}
\\ A_{000}^{(2)}\end{array}\right)_{\rm NR} \, ,\nonumber\\
A_{++-}^{(2)}|_{\rm R} &=&  \left(\,1\,+\,i\,\delta_2\,\right)A_{++-}^{(2)}
|_{\rm NR}\, ,\nonumber\\
\ea
with the matrices
\ba
T_c\,=\,\frac{1}{3}\,\left(\begin{array}{cc}1&1\\1&-2 \end{array}\right)
\,,\hspace{2 cm} 
T_n\,=\,\frac{1}{3}\,\left(\begin{array}{cc}0&1\\-3&1 \end{array}\right)
\ea
projecting the final state with $I=1$ into the symmetric--non-symmetric 
basis \cite{DIP91}. The subscript R (NR) means that the re-scattering
effects  have (not) been included. 
In these definitions the matrix $\mathbb{R}$, $\delta_2$ 
and the amplitudes $A^{(i)}$ depend on $s_1$, $s_2$ and $s_3$. 

Up to linear terms in $y$, equation (\ref{Rescat})  is equivalent to
\ba \label{Rdelta2def}
\left(\begin{array}{c}-\alpha_1+\frac{1}{2}\alpha_3
\\-\beta_1+\frac{1}{2}\beta_3 \end{array}\right)_{\rm R} &=& 
\left(\,{\mathbb{I}}+
i\,\mathbb{R}\,\right)\left(\begin{array}{c}-\alpha_1+\frac{1}{2}\alpha_3
\\-\beta_1+\frac{1}{2}\beta_3 \end{array}\right)_{\rm NR}\,, \nonumber\\
\left(\begin{array}{c}\alpha_1+\alpha_3
\\ \beta_1+\beta_3 \end{array}\right)_{\rm R} &=& 
\left(\,{\mathbb{I}}+i\,\mathbb{R}\,\right)
\left(\begin{array}{c}\alpha_1+\alpha_3
\\ \beta_1+\beta_3 \end{array}\right)_{\rm NR} \,,\nonumber\\
\gamma_{3,{\rm R}} &=&  
\left(\,1\,+\,i\,\delta_2\,\right)\gamma_{3,{\rm NR}}\,.
\nonumber\\
\ea
Here, the matrix $\mathbb{R}$ and
 $\delta_2$ are functions of the meson masses and the pion decay coupling. 
At lowest order in the chiral counting they are  given by
\ba
\mathbb{R}^{\rm LO}
 &=&\frac{1}{32 \pi F_0^2}\sqrt{\frac{\mkd-9\mpd}{\mkd+3\mpd}}
\left(\begin{array}{cc}\frac{1}{3}(9\mpd+2\mkd)&0\\
\frac{1}{3}\mkd&-5\frac{\mpd(m_K^4-27m_{\pi}^4)}{(\mkd+3\mpd)(\mkd-9\mpd)}
\end{array}\right)
\ea
and
\be
\delta_2^{\rm LO} \,=\,-\frac{1}{96\pi F_0^2}\,\mkd\, 
\sqrt{\frac{\mkd-9\mpd}{\mkd+3\mpd}}
\ee
in agreement with \cite{DIPP94}.

If we substitute the values of the masses and the coupling constant 
$F_0$, we get
\be
\mathbb{R}^{\rm LO}=\left(\begin{array}{cc}0.136&0
\\0.050&-0.143\end{array}\right)
\,,\hspace{1.5cm}\delta_2^{\rm LO}\,=\,-0.050\,.
\ee

We have also obtained the phase $\delta_2^{\rm NLO}$ 
and two combinations of the  $\mathbb{R}^{\rm NLO}$ matrix elements 
at NLO  when including the dominant FSI from 
two-bubbles obtained as explained before. 
The determination  of all the elements of $\mathbb{R}^{\rm NLO}$ 
would require the  calculation of the FSI at NLO for all the 
amplitudes in (\ref{amp1}) --we only have done the charged Kaon decays.
 The analytical expressions for these NLO 
quantities are given in Appendix \ref{phasesNLO}. 
Numerically, we  get
\ba \label{RmatrixNLO}
\frac{\left(-\alpha_1+\frac{1}{2}\alpha_3\right)_{\rm R}}
{\left(-\alpha_1+\frac{1}{2}\alpha_3\right)_{\rm NR}}
\Bigg|^{\rm NLO}=1+i\,0.156
\,,\hspace{2 cm}\nonumber\\
\frac{\left(-\beta_1+\frac{1}{2}\beta_3\right)_{\rm R}}
{\left(-\beta_1+\frac{1}{2}\beta_3\right)_{\rm NR}}\Bigg|^
{\rm NLO}=1+i\,0.569
\,, \hspace{1.5cm} {\rm and}
\hspace{1.5cm}\delta_2^{\rm NLO}\,=\,-0.104\,.
\ea

\subsection{Results on the Asymmetries in the Slope $g$}

As we have seen 
in Section \ref{LOresults}, the electroweak contribution
to $\Delta g$ at LO proportional to $\im(e^2 G_E)$ is at most around 
10\% of the leading contribution proportional to $ G_8$
while $\re(e^2 G_E)$ generates a negligible contribution.  
We include in our results the NLO absorptive part of the
electroweak  amplitude which is proportional to
$\im (e^2 G_E)$. The rest of the electroweak amplitude is just used
in the estimate of the errors.\footnote{The expressions for the
order $e^2p^0$ and $e^2p^2$ contributions to all the decay $K \to 3 \pi$
amplitudes are in Appendix \ref{ZiK3pi}.}. 

In  order to study the NLO
effects in $g_{C(N)}$  and $\Delta g_{C(N)}$, 
it is convenient  to introduce 
\ba
\label{amp2}
\vert A(K^+ \rightarrow 3\,\pi) \vert ^2 &=& A^+_0 \,+\,y\,A^+_y
\,+\,\order(x,y^2)\,,\nonumber\\
\vert A(K^- \rightarrow 3\,\pi) \vert ^2 &=& A^-_0 \,+\,y\,A^-_y
\,+\,\order(x,y^2)\,,
\ea
so that
\ba \label{AmpDeltag}
&&\hspace{0.7 cm}g[K^{+(-)}\to 3\pi] \,=\, \frac{A^{+(-)}_y}{A^{+(-)}_0} 
\,,\nonumber\\
&&\Delta g \,=\, \frac{A^+_yA^-_0 - A^+_0 A^-_y}
{A^+_yA^-_0 + A^+_0A^-_y}\,.
\ea
Notice that the numerator and denominator in 
(\ref{AmpDeltag}) are not the same as the difference 
$g[K^+\to 3\pi]-g[K^-\to 3\pi]$ and the sum 
$g[K^+\to 3\pi]+g[K^-\to 3\pi]$ respectively.
At NLO, the sum 
$A^+_yA^-_0 + A^+_0A^-_y$ does not contain
the FSI at NLO since they are part of the
order $p^6$ contributions, i.e. of the next-to-next-to-leading
order effects for the real parts.
However,   the difference   $A^+_yA^-_0 - A^+_0 A^-_y$
is proportional to the imaginary part of the amplitudes, 
therefore to have it  at NLO we must take into 
account the FSI phases, i.e. we need to include the FSI at NLO only 
in the imaginary part. 

The analytical expressions of the 
functions $A^{+(-)}_0$ and $A^{+(-)}_y$ at 
NLO are collected for the charged and the neutral Kaon cases in Appendix 
\ref{Adeltag}. From these expressions, we get the following
 numerical results 
\ba
\label{eq:gNLOeff}
\Delta g_{C}^{\rm NLO}\simeq \left \lbrack 0.66 \, 
\im G_8 +4.33\, \im \widetilde K_2 -18.11\, \im \widetilde K_3 
 -  0.07\,\im (e^2 G_E) \right \rbrack \times 10^{-2}\, ,
\nonumber \\ 
\Delta g_{N}^{\rm NLO}\simeq  - \left \lbrack 0.04 \, \im G_8 
+ 3.69\, \im \widetilde K_2  
+ 26.29\, \im \widetilde K_3 + 0.05 \, \im (e^2 G_E) 
\right \rbrack \times 10^{-2}\, . \nonumber \\
\ea
Where we have used the values for $\re \widetilde K_i$
from the fit to CP-conserving $K\to 3 \pi$ amplitudes \cite{BDP03}.
The NLO counterterms $\im G_8$, $\im (e^2 G_E)$ and
 $\im \widetilde K_3$ are scale independent.
In (\ref{eq:gNLOeff}), we have fixed the remaining scale dependence from 
$\im \widetilde K_2$ at  $\mu=M_\rho$.  
For the only two unknown counterterms 
$\im \widetilde K_2$ and $\im \widetilde K_3$, 
we have made two estimates of their effects.
First, using (\ref{assum1}) 
as explained in Section \ref{inputs}.
 The other  estimate  of the effects of $\im \widetilde K_2$
and $\im \widetilde K_3$ is to put them to zero and to vary their
known scale dependence between 
$\mu= M_\rho$ and $ \mu = 1.5$ GeV. We include the induced variation
as a further uncertainty in our predictions.

Our final results for the slope $g$ asymmetries at NLO are
in Table \ref{tabNLO}.
\TABLE{
\label{tabNLO}
\begin{tabular}{||c|c|c|c|c||}\hline
&$ \Delta g_C^{ \rm NLO}(10^{-5})$ & 
$ \Delta \Gamma_C^{ \rm NLO}(10^{-6})$  &
$ \Delta g_N^{\rm NLO}(10^{-5})$
&$ \Delta \Gamma_N^{\rm NLO}(10^{-6})$  
\\\hline\hline
$\widetilde K_i(M_\rho)$ from Table \ref{tabKvalues}
& $-2.4\pm 1.2$& $\lbrack -11,9
\rbrack$&$1.1 \pm 0.7 $ & $\lbrack -9,11\rbrack$ \\
$\widetilde K_i(M_\rho)=0$& $-2.4\pm 1.3$& $1.0 \pm 0.7 $&
$0.9  \pm 0.5 $  & $4.0\pm  3.2 $ \\
\hline
\end{tabular}
\caption{CP-violating predictions for the slope $g$ and the
decay  rates $\Gamma$ at NLO in CHPT.
 The details of the calculation are  
in Section \ref{NLOresults}.  The inputs used for
$\im G_8$ and $\im (e^2 G_E)$ are in   
(\ref{gluonpenguin}) and (\ref{EMpenguin}), respectively.}}
The central values  are obtained with the input values in Table 
\ref{tabKvalues} and the uncertainty includes the uncertainties 
of $\im G_8$, $\im (e^2 G_E)$,  the uncertainties
of the counterterms  quoted in Table \ref{tabKvalues},
 the variation due to the scale explained above and the error due to
the electroweak corrections.

The contribution  of the order $p^4$ 
counterterms $\im \widetilde K_i$
to $\Delta g_C$ is around 25\% using the values in Table
\ref{tabKvalues} and  the dominant contribution 
 is  the term  proportional to  $\im G_8$.  
 For $\Delta g_N$  we find a much larger dependence
on the values of the $\im \widetilde K_i$. 
Of course, since $\im \widetilde K_i$ are unknown
these results should be taken just as order of magnitude results,
a factor of two  or three could not be unreasonable
for $\Delta g_C$ and $\Delta g_N$.
The contribution
of $\im (e^2 G_E)$ is smaller than a 10\% of the dominant one
for both $\Delta g_C$ and $\Delta g_N$.

\subsection{Results on the Asymmetries in the Decay Rates}

We also only include NLO absorptive 
electroweak effects proportional to $\im (e^2 G_E)$
 for the same reasons explained in the previous subsection.
The analytical functions $|A_{C(N)}^{NLO}|^2$ and  
$\Delta |A_{C(N)}^{NLO}|^2$ 
needed to obtain the asymmetries in (\ref{eq:dGLO}) 
at NLO are given in (\ref{eqANLO}). 
Also  as explained in the previous subsection,
one should consistently not include 
FSI at NLO, which are  order $p^6$,
in the squared amplitudes $|A_{C(N)}^{NLO}|^2$  
since  they are part of the next-to-next-to-leading order
 corrections.
On the contrary, one has to include FSI at NLO
 in the differences $\Delta |A_{C(N)}^{NLO}|^2$ since 
these differences  are proportional to the FSI phases. 

The results obtained numerically from (\ref{eqANLO}) in terms of the 
imaginary  part of the counterterms are
\ba
\Delta \Gamma_{C}^{\rm NLO}\simeq \left \lbrack -2.8 \, 
\im G_8 + 49.2\, \im \widetilde K_2 + 103.6\, \im \widetilde K_3 
 +  0.2\,\im (e^2 G_E) \right \rbrack \times 10^{-3}\, ,
\nonumber \\ 
\Delta \Gamma_{N}^{\rm NLO}\simeq   \left \lbrack -3.1 \, \im G_8 
+45.7 \, \im \widetilde K_2  
+56.3 \, \im \widetilde K_3+0.12 \, \im (e^2 G_E) 
\right \rbrack \times 10^{-3}\, . \nonumber \\
\ea
In both cases the final value of the asymmetry 
is strongly dependent on the 
exact value of the $\im \widetilde K_i$ 
due  to large cancellations in the contribution
proportional to $\im G_8$. If we use the 
uncertainties quoted in Table \ref{tabNLO} 
for $\im \widetilde K_i$, the  induced errors in $\Delta \Gamma_{C}$
and $\Delta \Gamma_{N}$ are over 100\%.
In Table \ref{tabNLO}, we just quote therefore ranges
for the two decay rates CP-violating asymmetries.

\section{Comparison with Earlier Work and Conclusions}
\label{conclu}

\subsection{Comparison with Earlier Work}
The asymmetries $\Delta g_C$ and $\Delta g_N$ 
have been discussed in the literature before finding
conflicting results.
The rather large result 
\be
\left|\Delta g_C\right|\simeq \left| \Delta g_N \right| 
\simeq 140.0 \times 10^{-5}\ ,
\ee
was found in \cite{BBEL89}.

The upper bounds
\be
\left| \Delta g_C \right| \leq 0.7 \times 10^{-5} \, , 
\ee
at LO  and 
\be
\left| \Delta g_C \right| \leq 4.5  \times 10^{-5} \, , 
\ee
at NLO were found in \cite{DIP91}. The NLO bound was obtained
making plausible assumptions  since  no full NLO  result in CHPT was used.
 
In  \cite{IMP92}, 
\be
\Delta g_C  \simeq - 0.16 \times 10^{-5} \ ,
\ee 
was found at LO and 
\ba
\Delta g_C  \simeq - (0.23 \pm 0.06) \times 10^{-5} \, 
\, {\rm and} \, \, \Delta g_N  \simeq  (0.13 \pm 0.04) \times 10^{-5} 
\ea  
in \cite{MP95} also at LO.
 The authors of \cite{IMP92,MP95}  also made some estimate 
of the NLO corrections and arrived to the conclusion that they
  could increase their LO result up to one order of magnitude.
But again no full NLO calculation in CHPT was used.

The asymmetries  $\Delta \Gamma_C$ and $\Delta \Gamma_N$ have also
been discussed before and the results found were also in conflict
among them: 
\ba
\left|\Delta \Gamma_C\right|\simeq 31.0 \times 10^{-6} 
&\hspace*{0.5cm} {\rm and} \hspace*{0.5cm} 
& \left| \Delta \Gamma_N \right| 
\simeq 100.0 \times 10^{-6}\ ,
\ea
in \cite{BBEL89} ,
\be
\label{resIMP}
\Delta \Gamma_C  \simeq - 0.04 \times 10^{-6} \ ,
\ee 
in \cite{IMP92}, 
\ba
\label{resMP}
\Delta \Gamma_C  \simeq - (0.06 \pm 0.02) \times 10^{-6} 
\, \, {\rm and} \, \, 
\Delta \Gamma_N  \simeq (0.24 \pm 0.08) \times 10^{-6} 
\ea 
in \cite{MP95}, and  
\be
\label{resSHA}
\Delta \Gamma_C  \simeq - 1.0 \times 10^{-6} \ ,
\ee
in \cite{SHA93} --where we have used $\sin (\delta_{SM}) \simeq 0.85$
\cite{PDG02}.
The result in \cite{BBEL89}
 was claimed to be at one--loop, however
they did not use CHPT fully at one--loop. 
 We find, in general, that the results in \cite{BBEL89}
are overestimated as already pointed out in
 \cite{DIP91,IMP92,MP95,SHA03,SHA93}. See \cite{DIP91}
where some explanations for this large discrepancy are 
discussed.

The results in \cite{DIP91,IMP92,MP95} were reviewed
in \cite{DI96}.
They used factorizable values for $\im G_8$ and $\im G_E$,
i.e. the couplings in (\ref{largeNccouplings}), 
therefore their results have to be compared with the first row
in Table \ref{tabLO}.  
The reason of the difference between their results and ours 
is due mainly to the fact that these authors obtain the value 
of $\re G_8$ using the experimental value
for the isospin I=0 $K\to \pi \pi$
amplitude $\re a_0$. This amplitude $\re a_0$
contains  large higher order in CHPT
corrections. Corrections of similar size occur also 
in $\im a_0$ when considered at all orders.
However the authors used analytic LO formulas
for $\im a_0$ as well as for the imaginary parts of $K \to 3\pi$
amplitudes. This asymmetric procedure of considering the
real parts of the amplitudes experimentally and
the imaginary parts  analytically just at LO 
leads to a value for $\re G_8$  which is overestimated.
Therefore the CP violating asymmetries at LO
are underestimated in \cite{DIP91,IMP92,MP95}. 
The same comments apply to the predictions
of $\varepsilon_K'$ in those references as emphasized in \cite{PPI01}.
Our result in Tables \ref{tabLO} and \ref{tabNLO}
fulfill numerically the upper bound found in \cite{DIP91}
for $\Delta g_C$ at NLO but not the upper bound found
there at LO because of the same reason explained above.

The results in \cite{SHA93} were obtained at NLO using the
linear $\sigma$-model.
Recently, there was an update of those results in \cite{SHA03}:
\be
\label{7.9}
\Delta g_C  \simeq - (3.4 \pm 0.6)\times 10^{-5} \ ,
\ee 
at LO and
\be
\label{7.10}
\Delta g_C  \simeq - (4.2 \pm 0.8)\times 10^{-5} \ ,
\ee 
at NLO in the linear $\sigma$-model.  It is , however,  unclear from 
the text, the values used for 
the gluonic and the  electroweak penguins matrix elements
to get those results.
Though the LO result in (\ref{7.9}) agrees numerically
 with our result in Table \ref{tabLO}, we   do not agree analytically
with the results in \cite{SHA03} when the author
says that the electroweak penguins contribution at LO is as much as
34\% of the gluonic penguins contribution.
 We find that the electroweak
penguin contribution is one order of magnitude suppressed
with respect to the gluonic one.
\subsection{Conclusions}
We have performed the first full analysis
at NLO in CHPT of the CP-violating asymmetries
in the slope $g$ and the decay rate $\Gamma$ for the
disintegration  of charged Kaons into three pions.
 We have done the full order $p^4$ calculation
for $K \to  3 \pi$ and completely agree with the recent results
in \cite{BDP03}.
 To give the CP-asymmetries at NLO, one needs the FSI phases at NLO
also, i.e. at two loops.
We have calculated the dominant two-bubble contributions
using the optical theorem and the known one-loop and tree level
results in Appendix \ref{FSI6} as explained in
Section \ref{6.1}.  Due to the small phase
 space available for the 
re-scattering effects of the final tree pions one expects the rest 
of the FSI  to be  very suppressed.
We have included this contribution in our final numbers.
As a byproduct, we have  predicted the isospin I=2  FSI phase at NLO
and two combinations of matrix elements of
the isospin I=1 FSI re-scattering matrix $\mathbb{R}$ at NLO.
They can be found numerically  in 
Section \ref{6.1} and analytically in Appendix \ref{phasesNLO}.
We have given  analytical expressions for all the results in the 
Appendices \ref{Amplitudes}, \ref{Adeltag}, \ref{ANLO},
and  \ref{FSI6}.  

 Our final results at LO can be found in Table \ref{tabLO}
and at NLO in Table \ref{tabNLO}. If we use the counterterms
in Table \ref{tabKvalues}, we find  NLO corrections of 
the expected size, i.e. around 20\%,    for $\Delta g_C$
and $\Delta g_N$.
With those values for the NLO counterterms, the CP-violating 
asymmetry $\Delta g_C$ is dominated by  the value of $\im G_8$
while the rest of the CP-violating asymmetries studied here, 
namely,  $\Delta g_N$, $\Delta \Gamma_C$ and $\Delta \Gamma_N$, 
are dominated by the value of $\im \widetilde K_2$
and $\im \widetilde K_3$. 

Of course, our results in Table \ref{tabNLO} 
depend on the size of
$\im \widetilde K_2$ and $\im \widetilde K_3$.
If their values are within a factor two to three
the ones in Table \ref{tabKvalues} then the central
value of $\Delta g_C$
changes within the quoted uncertainties for it,
while the central value for $\Delta g_N$ doubles.
 The asymmetries in the decay rates
$\Delta \Gamma_C$ and $\Delta \Gamma_N$ can change even sign
if we vary $\im \widetilde K_2$ and $\im \widetilde K_3$ within
the uncertainties quoted in Table \ref{tabKvalues}.
Therefore, we have presented for them just ranges.

We partially disagree with
references \cite{DIP91,MP95,DI96} when  the authors claim that one 
could expect one order of magnitude enhancement at NLO
in all the asymmetries studied here. 
We find that for $\Delta g_C$ and $\Delta g_N$ the NLO corrections
are of the order of 20\% to 30\% . Only $\Delta \Gamma_C$
and $\Delta \Gamma_N$  can vary of one order
of magnitude and even change sign depending on the value
of $\im \widetilde K_2$ and $\im \widetilde K_3$.
We also find that $\Delta g_C$ can be as large as $-4 \times
10^{-5}$  both at LO and NLO while in the
conclusions of \cite{DIP91,MP95,DI96}  it was
claimed that any of these asymmetries could not exceed $10^{-5}$
within the Standard Model.
 
In Section \ref{5.2}, we found 
that making the cut proposed in \cite{AVI81,GRW86}
for the energy of the pion with charge opposite to the
decaying Kaon, there is one order of magnitude enhancement
for $\Delta \Gamma_C$
in agreement with the claims in those references. 
This result is however  valid for our LO calculation. 
It remains unclear whether the cut can provide a real
advantage at NLO  since in this case the cancellation among the various 
counterterm contributions can mask the effect. In addition, 
it remains to see how feasible is  to perform this cut experimentally. 
We do not find this enhancement for $\Delta \Gamma_N$. 

 The measurement of these CP-violating asymmetries
 by NA48 at CERN and/or  by KLOE at Frascati 
and/or elsewhere at the level of $10^{-4}$ to 
$10^{-5}$ will  be extremely interesting
for many reasons.
 The combined analysis of all four CP-violating asymmetries
$\Delta g_C$, $\Delta g_N$, $\Delta \Gamma_C$ and 
$\Delta \Gamma_N$ can allow to  obtain more information
on the  values of the presently poorly known $\im G_8$,
and the  unknown $\im \widetilde K_2$ and $\im \widetilde K_3$.
Due to the different dependence on these parameters, 
if the measurement  is  good enough, 
one can  try to fix $\im \widetilde K_2$ and 
$\im \widetilde K_3$ from the measurement of the asymmetries
$\Delta g_N$, $\Delta \Gamma_C$ and $\Delta \Gamma_N$
which are dominated by  the order $p^4$ counterterms and
use them to predict more accurately $\Delta g_C$.

 The large dependence of the  asymmetry $\Delta g_{C}$ 
 of $\im G_8$ at NLO can also be used as 
 consistency check between the theoretical 
predictions for  $\Delta g_C$ and  for the CP-violating
parameter $\varepsilon_K'$.  Any prediction
for $\varepsilon_K'$ has to be also able to predict the CP-violating
asymmetries discussed here.  In particular, 
the measurement of $\Delta g_C$ may  also
 shed light on a possible large  value for $\im G_8$ 
as found in calculations
at NLO in $1/N_c$ --see for instance \cite{BP00,HPR03,HKPS00}.

Moreover, it seems that some  models beyond the Standard Model can reach
values not much larger than $1\times 10^{-4}$ for the 
CP-violating  asymmetries, see for instance \cite{DIM00}. 
Our results can help to distinguish  new physics effects 
from the Standard Model ones in these  observables
and unveil beyond the Standard Model physics.

\section*{Acknowledgments}

We want to thank Hans Bijnens for useful comments and 
reading the manuscript and Toni Pich for encouragement and discussions.
I.S. acknowledges also conversations with Gilberto Colangelo. 
This work has been supported in part by the European Union
RTN Grant No HPRN-CT-2002-00311 (EURIDICE).
 E.G. is indebted to MECD (Spain) for a F.P.U. fellowship.
The work of E.G. and J.P. has been supported in part by MCYT (Spain)
 under Grants No. FPA2000-1558 and HF2001-0116 and  
by Junta de Andaluc\'{\i}a Grant No. FQM-101. 
The work of I.S. has been supported in part 
by the Swiss National Science Foundation and RTN, BBW-Contract 
No. 01.0357. 

\section*{Appendices}

\appendix
\section{$\Delta S=1$ Chiral Lagrangian}
\label{LagNLO}

 At next to leading order, the SU(3) $\times$ SU(3)
 chiral Lagrangian describing $K\to 3 \pi$ decays is given by
\ba
\label{8deltaS1}
{\cal L}^{(4)}_{|\Delta S|=1}&=&
 C F_0^2  \re{G_8} \, 
\left\{ N_1 {\cal O}_1^8 + N_2 {\cal O}_2^8 + N_3 {\cal O}_3^8 +
 N_4 {\cal O}_4^8 + N_5 {\cal O}_5^8 + N_6 {\cal O}_6^8 + 
N_7 {\cal O}_7^8 \right. \nonumber \\ 
&+& \left. N_8 {\cal O}_8^8 +  N_9 {\cal O}_9^8 +
N_{10}{\cal O}_{10}^8 +  N_{11} {\cal O}_{11}^8 
+ N_{12} {\cal O}_{12}^8 + N_{13} {\cal O}_{13}^8 \right\}  + {\rm h.c.}
\ea
for the octet part \cite{KMW90,EF91,EKW93},
\ba
\label{27deltaS1}
{\cal L}^{(4)}_{|\Delta S|=1}&=&
 C F_0^2 G_{27} \, 
\left\{ D_1 {\cal O}_1^{27} + D_2 {\cal O}_2^{27} + 
D_4 {\cal O}_4^{27} + D_5 {\cal O}_5^{27}+ D_6 {\cal O}_6^{27}  +
D_7 {\cal O}_7^{27}  \right. \nonumber \\ 
&+&\left. D_{26} {\cal O}_{26}^{27} +  D_{27} {\cal O}_{27}^{27}  
+ D_{28}{\cal O}_{28}^{27} + D_{29} {\cal O}_{29}^{27}  
D_{30} {\cal O}_{30}^{27} + D_{31} {\cal O}_{31}^{27} \right\} 
 + {\rm h.c.}
\ea
 for the 27-plet part \cite{KMW90,EF91} and 
\ba
\label{EMdeltaS1}
{\cal L}^{(4)}_{|\Delta S|=1}&=&
 C e^2 F_0^4 \re{G_{8}} \, 
\left\{ Z_1 {\cal O}_1^{EW} + Z_2 {\cal O}_2^{EW} + 
Z_3 {\cal O}_3^{EW} + Z_4 {\cal O}_4^{EW} + Z_5 {\cal O}_5^{EW}  
Z_6 {\cal O}_6^{EW} \right.  \nonumber \\ 
&+& Z_{7} {\cal O}_{7}^{EW} + Z_{8} {\cal O}_{8}^{EW}  
+ Z_{9}{\cal O}_{9}^{EW} + Z_{10} {\cal O}_{10}^{EW}  \nonumber \\
&+& \left. Z_{11} {\cal O}_{11}^{EW} + Z_{12} {\cal O}_{12}^{EW} 
+ Z_{13} {\cal O}_{13}^{EW}  + Z_{14} {\cal O}_{14}^{EW} \right\} 
 + {\rm h.c.}
\ea
 for the electroweak part with the 
dominant octet structure \cite{EIMNP00}. 

The octet operators are
\ba
{\cal O}_1^8 = \tr \left( \Delta_{32} u_\mu u^\mu u_\nu u^\nu \right), 
\; &
{\cal O}_2^8  = \tr \left( \Delta_{32} u_\mu u_\nu u^\nu u^\mu \right),
\; \nonumber \\
{\cal O}_3^8 = \tr \left( \Delta_{32} u_\mu u_\nu \right)
\tr \left( u^\mu u^\nu \right), \; &
   {\cal O}_4^8 = \tr \left( \Delta_{32} u_\mu \right) 
\tr \left( u_\nu u^\mu u^\nu \right),  \;  \nonumber \\
{\cal O}_5^8 = \tr \left( \Delta_{32} \left( \chi_+ u_\mu u^\mu
 + u_\mu u^\mu \chi_+\right) \right) , \;  &
  {\cal O}_6^8 = \tr \left( \Delta_{32} u_\mu \right) 
\tr \left( u^\mu \chi_+ \right), \; \nonumber \\ 
{\cal O}_7^8 = \tr \left( \Delta_{32} \chi_+\right) 
\tr \left( u_\mu u^\mu \right), \; & 
  {\cal O}_8^8 = \tr \left( \Delta_{32} u_\mu u^\mu \right) 
\tr \left( \chi_+ \right), \; \nonumber \\
{\cal O}_9^8 = \tr \left( \Delta_{32} \left( \chi_- u_\mu u^\mu
 - u_\mu u^\mu \chi_-\right) \right) , \; & 
  {\cal O}_{10}^8 = \tr \left( \Delta_{32} \chi_+ \chi_+\right), 
 \; \nonumber \\ 
{\cal O}_{11}^8 = \tr \left( \Delta_{32}  \chi_+\right) 
\tr \left( \chi_+ \right), \; & 
  {\cal O}_{12}^8 = \tr \left( \Delta_{32} \chi_- \chi_-\right), 
 \; \nonumber \\ 
{\cal O}_{13}^8 = \tr \left( \Delta_{32} \chi_-\right)
\tr \left(\chi_-\right)\, .  & 
\ea

 The 27-plet operators are
\ba
{\cal O}_1^{27}= t^{ij,kl} \tr \left( \Delta_{ij} \chi_+\right) 
\tr\left( \Delta_{kl} \chi_+\right), \; \nonumber \\
{\cal O}_2^{27}= t^{ij,kl} \tr \left( \Delta_{ij} \chi_-\right) 
\tr\left( \Delta_{kl} \chi_-\right), \; \nonumber \\
{\cal O}_4^{27}= t^{ij,kl} \tr \left( \Delta_{ij} u_\mu\right) 
\tr\left( \Delta_{kl}
\left(  u^\mu \chi_+ + \chi_+ u^\mu \right) \right), \; \nonumber \\
{\cal O}_5^{27}= t^{ij,kl} \tr \left( \Delta_{ij} u_\mu\right) 
\tr\left( \Delta_{kl}
\left(u^\mu \chi_- - \chi_- u^\mu \right) \right), 
\; \nonumber \\
{\cal O}_6^{27}= t^{ij,kl} \tr \left( \Delta_{ij} \chi_+\right) 
\tr\left( \Delta_{kl} u^\mu u_\mu \right), \; \nonumber \\
{\cal O}_7^{27}= t^{ij,kl} \tr \left( \Delta_{ij} u_\mu\right) 
\tr\left( \Delta_{kl} u^\mu \right) \tr \left( \chi_+ \right), \; 
\nonumber \\
{\cal O}_{26}^{27}= t^{ij,kl} \tr \left( \Delta_{ij} u^\mu u_\mu\right) 
\tr\left( \Delta_{kl} u^\nu u_\nu \right) , \; \nonumber \\
{\cal O}_{27}^{27}= t^{ij,kl} \tr \left( \Delta_{ij} \left(u_\mu u_\nu
+ u_\nu u_\mu \right) \right) \tr \left( \Delta_{kl} \left(u^\mu u^\nu
+ u^\nu u^\mu \right) \right), \; \nonumber \\
{\cal O}_{28}^{27}= t^{ij,kl} \tr \left( \Delta_{ij} \left(u_\mu u_\nu
- u_\nu u_\mu \right) \right) \tr \left( \Delta_{kl} \left(u^\mu u^\nu
- u^\nu u^\mu \right) \right), \; \nonumber \\
{\cal O}_{29}^{27}= t^{ij,kl} \tr \left( \Delta_{ij} u_\mu \right) 
\tr \left( \Delta_{kl} u_\nu u^\mu u^\nu\right), \; \nonumber \\
{\cal O}_{30}^{27}= t^{ij,kl} \tr \left( \Delta_{ij} u_\mu \right) 
\tr \left( \Delta_{kl} \left(u^\mu u_\nu u^\nu + 
u_\nu u^\nu u^\mu \right)\right), \; \nonumber \\
{\cal O}_{29}^{27}= t^{ij,kl} \tr \left( \Delta_{ij} u_\mu \right) 
\tr \left( \Delta_{kl} u^\mu \right) \tr \left( u_\nu u^\nu \right).
\ea

 The dominant octet electroweak operators are
\ba
{\cal O}_1^{EW}= \tr \left( \Delta_{32} 
\left\{ u^\dagger Q u ,\chi_+ \right\}\right) , \; 
&  {\cal O}_2^{EW}= \tr \left( \Delta_{32} 
u^\dagger Q u \right) \tr \left( \chi_+\right), \nonumber \\
{\cal O}_3^{EW}= \tr \left( \Delta_{32} 
u^\dagger Q u \right) \tr \left( \chi_+ u^\dagger Q u \right), 
\; &  {\cal O}_4^{EW}= \tr \left( \Delta_{32} \chi_+\right) 
\tr \left(  Q U^\dagger Q U \right), \nonumber \\ 
{\cal O}_5^{EW}= \tr \left( \Delta_{32} u^\mu u_\mu \right) , \; 
&  {\cal O}_6^{EW}=  \tr \left( \Delta_{32} 
\left\{ u^\dagger Q u , u^\mu u_\mu \right\}\right) , \; \nonumber \\
{\cal O}_7^{EW}= \tr \left( \Delta_{32} u^\mu u_\mu \right) 
\tr \left( Q U^\dagger Q U \right) , \; 
&  {\cal O}_8^{EW}=  \tr \left( \Delta_{32}u^\mu \right) 
\tr \left(Q u^\dagger  u_\mu u \right), \; \nonumber \\
{\cal O}_{9}^{EW}=  \tr \left( \Delta_{32}u^\mu \right) 
\tr \left(Q u  u_\mu u^\dagger \right), \; &
{\cal O}_{10}^{EW}=  \tr \left( \Delta_{32}u^\mu \right) 
\tr \left(\left\{u Q u^\dagger,   u^\dagger  Q u \right\} 
u_\mu \right), \; \nonumber \\
{\cal O}_{11}^{EW}=  
\tr \left(\Delta_{32} \left\{u^\dagger  Q u , u^\mu \right\} \right)
\tr \left( u Q u^\dagger u_\mu \right) , \; & 
{\cal O}_{12}^{EW}=  
\tr \left(\Delta_{32} \left\{u^\dagger  Q u , u^\mu \right\} \right)
\tr \left( u^\dagger  Q u  u_\mu \right) , \; \nonumber \\
{\cal O}_{13}^{EW}=  
\tr \left(\Delta_{32} u^\dagger  Q u  \right)
\tr \left( u^\mu u_\mu \right) , \; & 
{\cal O}_{14}^{EW}=  
\tr \left(\Delta_{32} u^\dagger  Q u  \right)
\tr \left( u^\dagger  Q u  u_\mu u^\mu\right) . \nonumber \\
\ea

We have done the NLO calculation in the presence of strong interactions
which at LO order are described by \cite{WEI79,GL}
\ba \label{EMp2}
{\cal L}^{(2)} &=& \frac{F_0^2}{4}
\left[ \tr \left( u_\mu u^\mu \right) + \tr \left(\chi_+ \right) \right];
 \ea
at NLO, the SU(3) $\times$ SU(3)
 strong chiral Lagrangian needed in $K\to 3 \pi$ decays is given by
\ba
{\cal L}^{(4)} &=& 
L_1 \tr \left( u_\mu u^\mu \right)^2 
+L_2 \tr \left( u^\mu u^\nu \right) \tr \left( u_\mu u^\nu \right)
+L_3 \tr \left( u^\mu u_\mu u^\nu u_\nu\right)
+L_4 \tr \left( u^\mu u_\mu \right) \tr \left( \chi_+ \right)
\nonumber \\
&+& L_5 \tr \left( u^\mu u_\mu \chi_+ \right)
+L_6 \tr \left( \chi_+ \right) \tr \left (\chi_+ \right)
+L_7 \tr \left( \chi_-\right) \tr \left( \chi_-\right)
+\frac{1}{2} L_8 \tr \left( \chi_+ \chi_+ + \chi_- \chi_- \right) \, .
\nonumber \\
\ea

\section{$K\to 3 \pi$ Amplitudes at NLO}
\label{Amplitudes}

A general way of writing the decay amplitude for $K^+\rightarrow 3\pi$ at 
NLO including FSI effects also at NLO is
\ba \label{genamplitude}
 A(K^+\rightarrow 3\pi)\,(s_1,s_2,s_3)= G_8\,a_{8}(s_1,s_2,s_3)\,
+\,G_{27}\,a_{27}(s_1,s_2,s_3)\,+\,e^2 G_E\,a_{E}(s_1,s_2,s_3)\,
\hspace{0.5 cm}&&\nonumber\\
+\,F^{(4)}(\widetilde K_i,Z_i,s_1,s_2,s_3)
\,+\,i\,F^{(6)}(\widetilde K_i,Z_i,s_1,s_2,s_3)\,.
\hspace{2 cm}&&
\ea
While for the corresponding CP conjugate the amplitude is
\ba
A(K^-\rightarrow 3\pi)\,(s_1,s_2,s_3)= G_8^*\,a_{8}(s_1,s_2,s_3)\,
+\,G_{27}\,a_{27}(s_1,s_2,s_3)\,+\, e^2 G_E^*\,a_{E}(s_1,s_2,s_3)\,
\hspace{0.5 cm}&&\nonumber\\
+\,F^{(4)}(\widetilde K_i^*,Z_i^*,s_1,s_2,s_3) 
\,+\,i\,F^{(6)}(\widetilde K_i^*,Z_i^*,s_1,s_2,s_3)\,.
\hspace{2 cm}&&
\ea
The energies $s_i$ are defined in Section \ref{notat},
 the $\widetilde K_i$  and $Z_i$ 
are counterterms appearing at  $\order(p^4)$
 and $\order(e^2 p^2)$ respectively,
 see Table \ref{tabKdef} and (\ref{EMdeltaS1}) for definitions. The functions 
$F^{(4)}(s_1,s_2,s_3)$ and $F^{(6)}(s_1,s_2,s_3)$ are
\ba \label{F4F6def}
F^{(4)}(\widetilde K_i,Z_i,s_1,s_2,s_3)&=&\sum_{i=1,11}H_i^{(4)}(s_1,s_2,s_3)
\widetilde K_i
+\sum_{i=1,14}J_i^{(4)}(s_1,s_2,s_3)Z_i \,,\nonumber\\
F^{(6)}(\widetilde K_i,Z_i,s_1,s_2,s_3)
&=& \sum_{i=1,11}H_i^{(6)}(s_1,s_2,s_3)
\widetilde K_i
+ \sum_{i=1,14}J_i^{(6)}(s_1,s_2,s_3)Z_i \,.
\ea

The complex functions $a_i$ can be written in terms of real functions 
as
\ba \label{functionsNLO}
a_i\,(s_1,s_2,s_3)\,&=&\,B_i(s_1,s_2,s_3)\,+\,i\,C_i(s_1,s_2,s_3)
\ea
$B_i(s_1,s_2,s_3)$ and $C_i(s_1,s_2,s_3)$ are real functions 
corresponding to the dispersive and absorptive amplitudes respectively and 
admit a CHPT expansion
\ba \label{amplitudesNLO}
B_i(s_1,s_2,s_3) &=& B_i^{(2)}(s_1,s_2,s_3) + B_i^{(4)}(s_1,s_2,s_3)
+\order(p^6)\,,\nonumber\\
C_i(s_1,s_2,s_3) &=& C_i^{(4)}(s_1,s_2,s_3) + C_i^{(6)}(s_1,s_2,s_3)
+\order(p^8)\, ,
\ea
where the superscript ${(2n)}$ indicates that the function is 
$\order (p^{2n})$ in CHPT. 

The functions $B_i^{(2)}$, $B_i^{(4)}$,  
$C_i^{(4)}$ and the part depending on $\widetilde K_i$ of $F^{(4)}$ in 
(\ref{genamplitude}) and (\ref{amplitudesNLO}) 
for $i=8,27$, which
 correspond to the CP-conserving amplitudes up to order 
$\order (p^4)$ and without electroweak corrections, that is,
\ba
A(K \to 3\pi ) &=& \re G_8 \left(B_8^{(2)}(s_1,s_2,s_3)+B_8^{(4)}(s_1,s_2,s_3)
+iC_8^{(4)}(s_1,s_2,s_3)\right)\nonumber\\
&&+ G_{27} \left(B_{27}^{(2)}(s_1,s_2,s_3)+B_{27}^{(4)}(s_1,s_2,s_3)
+iC_{27}^{(4)}(s_1,s_2,s_3)\right)\nonumber\\
&&+F^{(4)}(\re \widetilde K_i,s_1,s_2,s_3)\,
\ea
were obtained in \cite{BDP03}. 
We calculated these amplitudes for all the decays defined in (\ref{defdecays}) 
and got total agreement with \cite{BDP03}. 
The explicit expressions can be found there taking into account that the 
relation between the functions defined here and those used in \cite{BDP03} is, 
for the charged Kaon decays, 
\ba
&&M_{10}(s_3)+M_{11}(s_1)+M_{11}(s_2)+M_{12}(s_1)(s_2-s_3)+M_{12}(s_2)(s_1-s_3)
\nonumber\\
&&\hspace{1 cm}=\re G_8 \left(B_8^{(2)}+B_8^{(4)}+iC_8^{(4)}
\right)_{(++-)}\nonumber\\
&&\hspace{2 cm}+ G_{27} \left(B_{27}^{(2)}+B_{27}^{(4)}
+iC_{27}^{(4)}\right)_{(++-)}+F^{(4)}_{(++-)}\,,\nonumber\\
&&M_{7}(s_3)+M_{8}(s_1)+M_{8}(s_2)+M_{9}(s_1)(s_2-s_3)+M_{9}(s_2)(s_1-s_3)
\nonumber\\
&&\hspace{1 cm}=\re G_8 \left(B_8^{(2)}+B_8^{(4)}+iC_8^{(4)}
\right)_{(00+)}\nonumber\\
&&\hspace{2 cm}+ G_{27} \left(B_{27}^{(2)}+B_{27}^{(4)}
+iC_{27}^{(4)}\right)_{(00+)}+F^{(4)}_{(00+)}\, .
\ea

The functions $B_i^{(2)}$, $B_i^{(4)}$,  
$C_i^{(4)}$ and the part depending on $\widetilde K_i$ of $F^{(4)}$ in 
(\ref{genamplitude}) and (\ref{amplitudesNLO}) 
were calculated for $i=8,27$ in \cite{BDP03}. 
We calculated these quantities and got total agreement with \cite{BDP03},
 the explicit expressions can be found there. The functions 
$C_i^{(6)}(s_1,s_2,s_3)$ (for i=8,27) and $F^{(6)}$ are associated to FSI 
at NLO coming 
from two loops diagrams and are discussed in Appendix \ref{FSI6}. 

We have also calculated the contributions 
of order $e^2p^0$ and $e^2 p^2$ from the CHPT Lagrangian in (\ref{deltaS1}) 
and  (\ref{EMdeltaS1}) in presence of strong interactions for all the 
$K \to 3\pi$ transitions, that fix the functions $B_E^{(2)}$,  $B_E^{(4)}$, 
$C_E^{(4)}$ and $J_i^{(4)}$. The results are in the next subsection.

In order to calculate the asymmetries in the slope g defined in 
(\ref{gdefinition}) we need to expand these amplitudes in powers of the 
Dalitz plots variables $x$ and $y$, 
\ba
x\equiv \frac{s_1-s_2}{m_{\pi^+}^2} & {\rm and}&
\quad y \equiv \frac{s_3-s_0}{m_{\pi^+}^2}\,.
\ea
The notation we are going to use here is
\be \label{yexpansion}
G_i^{(2n)}(s_1,s_2,s_3)\,=\,G_{i,0}^{(2n)}\,+\,y\,G_{i,1}^{(2n)} 
\,+\,\order(x,y^2)\,;
\ee
where the function $G_i(s_1,s_2,s_3)$ can be any of the functions 
$B_i(s_1,s_2,s_3)$, $C_i(s_1,s_2,s_3)$ defined in (\ref{functionsNLO}) or 
$H_i(s_1,s_2,s_3)$, $J_i(s_1,s_2,s_3)$ in 
(\ref{F4F6def}). The coefficients $G_{i,0(1)}^{(2n)}$ are real quantities 
that depend on the masses $m_{\pi}^2$, $m_K^2$, 
the pion decay constant and the strong counterterm 
couplings of $\order (p^4)$, i.e., $L_i^r$.

\subsection{$\order(e^2 p^0)$  and $\order(e^2 p^2)$ Contributions}
\label{ZiK3pi}

 Here we give the order $e^2p^0$ and $e^2p^2$ contributions
to all the $K \to 3 \pi$ amplitudes without including  virtual photon loops
ones. We do not give the  order $e^2 p^2$ contributions
from one order $e^2 p^0$ $\Delta S=0$ vertex 
 when it is inserted   in the external Kaon and pion lines 
since these contributions are mainly 
responsible of the $m_{\pi^+}^2- m_{\pi^0}^2$
and $m_{K^+}^2-m_{K^0}^2$ mass  differences and we take 
physical masses for them.

  Electroweak interactions break in general isospin symmetry.
An isospin decomposition which is valid up to order $e^2p^2$ 
at least, is the following
\ba\label{AcBcdef}
A_{++-}(s_1,s_2,s_3)&=&
2 A_{c,+}(s_1,s_2,s_3)+B_{c}(s_1,s_2,s_3)+B_{t,+}(s_1,s_2,s_3)\, ,
 \nonumber \\
A_{00+}(s_1,s_2,s_3)&=&
A_{c,0}(s_1,s_2,s_3)-B_c(s_1,s_2,s_3)+B_{t,+}(s_1,s_2,s_3) \, ,
 \nonumber \\
A^{1}_{+-0}(s_1,s_2,s_3) &=& C_0^R(s_1,s_2,s_3)
+\frac{2}{3}\left[B_{t,0}^R(s_3,s_2,s_1)-
B_{t,0}^R(s_3,s_1,s_2)\right]\nonumber\\
 &&+A_{n,+}^I(s_1,s_2,s_3)-B_n^I(s_1,s_2,s_3)\, ,\nonumber\\
A^{2}_{+-0}(s_1,s_2,s_3) &=& C_0^I(s_1,s_2,s_3)
+\frac{2}{3}\left[B_{t,0}^I(s_3,s_2,s_1)-
B_{t,0}^I(s_3,s_1,s_2)\right]\nonumber\\
 &&+A_{n,+}^R(s_1,s_2,s_3)-B_{n}^R(s_1,s_2,s_3)\, ,\nonumber\\
A^{1}_{000}(s_1,s_2,s_3) &=&  3 A_{n,0}^I(s_1,s_2,s_3)\,,\nonumber\\
A^{2}_{000}(s_1,s_2,s_3) &=&  3 A_{n,0}^R(s_1,s_2,s_3)\,,
\ea
where $A_c$, $A_n$, $B_c$ and $B_n$ describe $I=1$ final states, 
$B_t$ the  $I=2$ one and $C_0$ the $I=0$ one. 
In (\ref{AcBcdef}) the superindex $R$  means that  only the real part of
 the counterterms appear in these functions 
and the superindex  $I$ that only the imaginary part is present.  
In the following we give the order $e^2p^2$ 
and $e^2p^0$ contributions to the 
functions defined in (\ref{AcBcdef}).

In order to write the formulas in a concise form
we define
\ba
C_{ew}&\equiv& - i\frac{3}{5}\frac{G_F}{\sqrt{2}}V_{ud}V_{us}^* 
\fr{F_0^6}{f_K f_{\pi}^3}e^2G_E\,, 
\nonumber \\
C_{ew}^R&\equiv& - i\frac{3}{5}\frac{G_F}{\sqrt{2}}V_{ud}V_{us}^*
\fr{F_0^6}{f_K f_{\pi}^3} \re(e^2G_E)\, , \nonumber \\
C_{ew}^I&\equiv& \frac{3}{5}\frac{G_F}{\sqrt{2}}V_{ud}V_{us}^* 
\fr{F_0^6}{f_K f_{\pi}^3}\im(e^2G_E)\,, \nonumber \\
\Delta&\equiv& m_K^2-m_\pi^2 \ ,\\
B^{LP}_k&\equiv&\overline{B}(m_L,m_P,s_k) \ ,\quad\quad
B^{LP}\equiv \overline{B}(m_L,m_P,s) \ , \\
D^{LP}_k&=&\overline{B}_1(m_L,m_P,s_k) \ ,\quad \quad
D^{LP}\equiv \overline{B}_1(m_L,m_P,s)  \ ,\\
\nu_L&\equiv&\fr{m_L^2}{16 \pi^2}\ln \left(\fr{m_L}{\nu}\right) \ ,
\ea
where $L,\ P=\pi,\ K,\ \eta$ and $k=1,2,3$. The functions 
$\overline{B}(m_L,m_P,s_k)$ and $\overline{B}_1(m_L,m_P,s_k)$ can be 
found,  e.g., in \cite{ABT00}. 
The constants $C_{ew}^{R(I)}$ must be used instead of 
$C_{ew}$ in the functions with superindex $R$ or $I$, respectively. 

\subsubsection{Electroweak Contributions at $\order(e^2 p^0)$}

\ba
A_{n,+}(s_1,s_2,s_3)=
A_{n,0}(s_1,s_2,s_3)&=& 0 \, ,\nonumber\\
A_{c,+}(s_1,s_2,s_3)=A_{c,0}(s_1,s_2,s_3)&=& -C_{ew} \, , \nonumber\\
B_n(s_1,s_2,s_3)&=& \frac{C_{ew}}{2\Delta}
(s_1+s_2-2s_3)\, ,\nonumber\\
B_{t,+}(s_1,s_2,s_3)=B_{t,0}(s_1,s_2,s_3)&=& -\frac{C_{ew}}{4\Delta}
(s_1+s_2-2s_3) \, ,\nonumber\\
B_c(s_1,s_2,s_3)&=& \frac{C_{ew}}{4\Delta}
(s_1+s_2-2s_3)\, ,\nonumber\\
C_0(s_1,s_2,s_3)&=& 0 \,.
\ea

\subsubsection{Electroweak Loop Contributions at $\order(e^2p^2)$}

  We checked  that the electroweak loops at order $e^2 p^2$ 
do not break the isospin symmetry and
\ba
A_{n,+}(s_1,s_2,s_3)=A_{n,0}(s_1,s_2,s_3)=A_n(s_1,s_2,s_3)\, , \nonumber \\
A_{c,+}(s_1,s_2,s_3)=A_{c,0}(s_1,s_2,s_3)=A_c(s_1,s_2,s_3)\, , \nonumber \\
B_{t,+}(s_1,s_2,s_3)=B_{t,0}(s_1,s_2,s_3)=B_t(s_1,s_2,s_3)\, . \nonumber \\
\ea
We get
\ba
A_n\vert_{Z_i=0} &=&\frac{C_{ew}}{f_{\pi}^2}\sum_{i=1,2,3} S_0(s_i) 
\,\,,  {\rm with}\nonumber\\
S_0(s) &=&
\frac{1}{36}\left( m_K^2 + 11\, m_\pi^2 \right) \,
    \left( \fr{1}{\Delta} + \frac{1}{s} \right) \,
    \left( \nu_K - \nu_\pi \right) -  \frac{1}{72 \Delta}
\Big[3 (3 s- m_K^2 - 3 m_\pi^2 ) 
\nonumber \\ \et
\times
 \left(s
    B^{KK} + 4 ( m_\pi^2 - s) B^{\pi\pi}\right) 
+\Big(\fr{\Delta^2}{s} (m_K^2+11m_\pi^2) -7m_K^4+ 26 m_K^2
\nonumber \\ \et \times m_\pi^2
+ 29 m_\pi^4
- s (5 m_K^2+67 m_\pi^2)+27 s^2\Big) B^{K\pi}\left.
+ 8   \Delta \left(\Delta + s\right) D^{K\pi}
\right] \, ,
\nonumber \\ 
\ea

\ba
A_c\vert_{Z_i=0}&=& 
\frac{C_{ew}}{f_\pi^2}\Big\{8\lbrack m_{\pi}^2
(-L_5^r+8L_6^r+4L_8^r)+\frac{2}{3}L_4^r(m_K^2-3m_{\pi}^2)\rbrack 
\nonumber\\ &&
+
\sum_{i=1,2,3} S_1(s_i)\Big\} \, , {\rm with} \nonumber\\
S_1(s)&=&\frac{1}{648\Delta }\Big[8 (19 m_K^2 - 6 m_\pi^2) \nu_\eta + 
4 (305
  m_K^2 - 318 m_\pi^2) (\nu_K + \nu_\pi)
\nonumber\\ \et\left.
 - \frac{2}{3 s}(27 \Delta^2 +s (217 m_K^2 - 435 m_\pi^2)) 
 (\nu_K -\nu_\pi) 
\right. -6 m_\pi^2 ( -17 m_K^2 
\nonumber\\ \et 
+ 5 m_\pi^2 
+ 9 s) B^{\eta\eta}
+162 \Delta s B^{KK} +
54 (m_\pi^4 - 10 m_\pi^2 s+ 3 s^2 
\nonumber\\ \et 
+ 3 m_K^2 (m_\pi^2+ s)) B^{\pi\pi} +2 (17 m_K^4 + 15 m_\pi^4 - m_K^2 (68 m_\pi^2 - 43 s) 
\nonumber\\ \et 
+ 17
m_\pi^2 s - 24 s^2) B^{\eta K} +\Big(\frac{9 \Delta^3}{ s} 
+2 (73 m_K^4-206 m_K^2 m_\pi^2
+25 m_\pi^4) 
\nonumber\\ \et
+3 (51
m_K^2
-19m_\pi^2) s
+16 s^2\Big) B^{K\pi}
-3
(-9 (5 m_K^4-6 m_K^2m_\pi^2
\nonumber\\ \et
+m_\pi^4)
+4 (m_K^2 -3m_\pi^2) s +5 s^2)
D^{\eta K}+( 261 
\times m_K^4 + 153 m_\pi^4
\nonumber\\ \et - 244 m_\pi^2 s + 275 s^2 
- m_K^2 (414 m_\pi^2 + 212 s))
D^{K\pi}\Big] \, ,
\ea

\ba
C_0|_{Z_i=0} &=& \frac{C_{ew}}{f_{\pi}^2}
\Big[ S_2(s_1,s_2,s_3)-S_2(s_2,s_1,s_3)\Big]\, ,
{\rm with} \nonumber\\ 
S_2(s_1,s_2,s_3) &=& \fr{s_1}{216\Delta}\Big[
6 (2 \nu_\eta-\nu_K-\nu_\pi)
+2\Big(3\Delta  (s_1 + s_2 - s_3)\fr{1}{s_1s_2} -\fr{1}{s_3}  (3\Delta 
 \nonumber\\ \et
 + 41 s_3)\Big) (\nu_K - \nu_\pi)
- 9 (s_3 -4 m_K^2 )B_3^{KK}-9 (5 m_K^2 - m_\pi^2 - s_3)   
\nonumber\\ \et
\!\! \times \!B_3^{\eta K}
 +\! \fr{3}{s_3}(\Delta\! +\!  s_3) (\Delta\! -\! 2 s_3)
B_3^{K\pi}-6 \Delta D_3^{\eta K}+6 (\Delta - 2 s_3)  D_3^{K\pi}\Big]
\nonumber\\ \et+
\fr{1}{216\Delta}\Big[
9 (4 m_K^2-s_1)
 (s_2-s_3)B_1^{KK}- \left(42 m_K^4 - 16 m_\pi^4 
\right. + m_K^2 
\nonumber\\ \et\left. 
 \times( 118  m_\pi^2- 81 s_3 - 61 s_1)
+ 
  m_\pi^2 (9 s_3 - 53 s_1) + 15 s_1 (3 s_3 + s_1)\right)
\nonumber\\ \et
\times B_1^{\eta K}\! +
\Big(\fr{\Delta}{s_1} + 1\Big) (3 \Delta (s_1+s_2 - s_3) - (8 m_K^2+22 m_\pi^2-3 s_3) s_1
\nonumber\\ \et
 + 44 s_1^2)B_1^{K\pi}
-\! 6 ( m_K^4 - 3 m_\pi^4 + 2 m_\pi^2 (s_3 - 5 s_1) + 2 m_K^2 (m_\pi^2 - s_3  
\nonumber\\ \et 
- 2 s_1)
+ s_1 (9 s_3 + 2 s_1)) D_1^{\eta K}+2 (3 \Delta (s_1+s_2 -  s_3) - (8 m_K^2  
\nonumber\\ \et
+ 22 m_\pi^2
- 3 s_3)  s_1 + 44 s_1^2)
D_1^{K\pi} \Big] \,,
\ea

\ba
B_t\vert_{Z_i=0}&=& 
\frac{ C_{ew}}{f_\pi^2}\Big\{ \frac{1}{\Delta}
\Big\lbrack L_3^r(s_3(s_1+s_2)-2s_1s_2)
 +2\,L_5^r(s_2+s_1-2s_3)m_{\pi}^2\Big\rbrack
\nonumber\\ &&
 +\left( S_3(s_1,s_2,s_3)
+S_4(s_1,s_2,s_3)+S_4(s_2,s_1,s_3)\right)\Big\}\, ,
{\rm with} \nonumber\\ 
S_3(s_1,s_2,s_3)&=&\fr{1}{432\Delta} \Big[\fr{-9}{(16\pi^2)}(5 m_K^4 + 39 m_\pi^4 + 4 m_K^2 (7
m_\pi^2 - 3 s_3) - 30 m_\pi^2 s_3  
\nonumber\\ \et
- 3 s_3^2
- 6 s_1 s_2)
-60 (s_1+s_2 - 2 s_3) (\nu_\eta + 13  (\nu_K + \nu_\pi)) 
\nonumber\\ \et
- \fr{2}{\Delta} \Big( \fr{108}{s_3} \Delta^2 m_\pi^2 - 31 m_K^4 -
26 m_K^2 m_\pi^2 +489 m_\pi^4 + 
3 (49 m_K^2 
\nonumber\\ \et
- 157 m_\pi^2) s_3  - 9 \Delta^2 
     (m_K^2 + 9 m_\pi^2   
- 2 s_3) \Big(\fr{1}{s_1} + \fr{1}{s_2}\Big) \Big)
(\nu_K - \nu_\pi)
\nonumber\\ \et-
( 54 (s_1+s_2 -  2 s_3) (2 m_\pi^2 - s_3) B_3^{\pi\pi}+2   (17 m_K^4
+ 15 m_\pi^4 
\nonumber\\ \et
-68m_\pi^2 m_K^2 + (17 m_\pi^2+43 m_K^2) s_3- 24 s_3^2 ) B_3^{\eta K} 
- \Big(\fr{108}{s_3} \Delta^2 m_\pi^2 
\nonumber\\ \et\!
 + ( 61 m_K^4 + 142 m_K^2 m_\pi^2 + 13 m_\pi^4) \! -\! 6 (31 m_K^2 + 51 m_\pi^2) s_3 + 
    185 s_3^2\Big) 
\nonumber\\ \et \! \times B_3^{K\pi}
\! +  
    3 ( 45 m_K^4 
+ 9 m_\pi^4 -54 m_K^2 m_\pi^2+ 4 (3 m_\pi^2-m_K^2) s_3 -\! 5 s_3^2 ) 
\nonumber\\ \et
\! \times D_3^{\eta K}+ (189 m_K^4 + 81 m_\pi^4 -270 m_K^2 m_\pi^2 
\nonumber\\ \et + (404
m_\pi^2-284 m_K^2)
 s_3 
+ 35 s_3^2 )
D_3^{K \pi}\Big)\Big] \, ,
\\
S_4(s_1,s_2,s_3)&=&\frac{1}{432 \Delta}\Big[ 
   54 (9 m_\pi^4 + m_K^2 (3 m_\pi^2 - s_1)+(s_1 - 4 m_\pi^2) (s_3 + 2 s_1) ) B_1^{\pi\pi}
\nonumber\\ \et
+(44 m_K^4 + 15 m_\pi^4 +m_K^2 (13 m_\pi^2 - 54 s_3 - 11 s_1) - 64 m_\pi^2 s_1 + 3 s_1
\nonumber\\ \et
 \times (18 s_3 + s_1)) B_1^{\eta K} - 
  \Big( 9 \fr{\Delta^2}{s_1}(m_K^2 + 9 m_\pi^2 - 2 s_3) - m_K^4  + 17 m_K^2 m_\pi^2
\nonumber\\ \et
- 16 m_\pi^4 + 9 (5 m_K^2  + m_\pi^2) s_3 - 3 (25 m_K^2 + 54 m_\pi^2 - 3 s_3) s_1+ 97 s_1^2\Big)
\nonumber\\ \et
 \times  B_1^{K\pi}- (-9 \Delta (8 m_K^2 - s_3)+ 3 (17 m_K^2  +33 m_\pi^2 - 27 s_3) s_1 - 33 s_1^2) 
\nonumber\\ \et
\times D_1^{\eta K}
- (-9 \Delta(10 m_K^2 - 6 m_\pi^2 + s_3)+ 
      5 (23 m_K^2 - 53 m_\pi^2 + 9 s_3) s_1 
\nonumber\\ \et 
+ 5 s_1^2)  D_1^{K\pi}\Big] \, ,
\ea

\ba
B_c\vert_{Z_i=0}&=& \frac{C_{ew}}{f_\pi^2}\Big\{\frac{1}{\Delta}
\Big\lbrack -L_3^r(s_3(s_1+s_2)-2s_1s_2)+\frac{2}{3}(8 \Delta L_4^r
-3m_\pi ^2L_5^r)
\nonumber\\ &&
\times (s_1+s_2-2s_3) 
\Big\rbrack
+ (S_5(s_1,s_2,s_3)+S_6(s_1,s_3)+S_6(s_2,s_3))\Big\}\, ,
{\rm with} \nonumber\\
\!
S_5(s_1,s_2,s_3)&=&\fr{1}{216\Delta} \Big[ \fr{9}{32\pi^2}
(5 m_K^4 + 39 m_\pi^4 + 28 m_K^2  m_\pi^2 - 6 s_3 (2 m_K^2 +5 m_\pi^2) 
\nonumber\\\et - 3(s_3^2 
+ 2 s_1 s_2))+2 (55 m_K^2+189 (m_\pi^2-s_3)) \nu_\eta+2 (175
m_K^2
\nonumber\\\et
+513 (m_\pi^2
-s_3))(\nu_K+\nu_\pi)
+\Big( 3\Delta (m_K^2 - 7 m_\pi^2 - 6 s_3)\Big(\fr{1}{s_1}+\fr{1}{s_2}\Big)
\nonumber\\\et
+\fr{12\Delta }{s_3}(m_K^2 
+ 8 m_\pi^2)
+\fr{1}{3\Delta}(163 m_K^4 + 554 m_K^2
m_\pi^2 + 579 m_\pi^4
- 243 
\nonumber\\\et\times (m_K^2 
+ 3 m_\pi^2)
 s_3)\Big)(\nu_K-\nu_\pi)+4 m_\pi^2 (-17 m_K^2 + 5 m_\pi^2 + 9 s_3) B_3^{\eta\eta}
\nonumber\\\et -108 s_3 \Delta B_3^{KK}
-9 (34 m_\pi^4 -42  m_K^2 m_\pi^2 +( 39 m_K^2 - 37 m_\pi^2)s_3 + 3 s_3^2)
\nonumber\\\et \times B_3^{\pi\pi}
-\fr{1}{3}(17 m_K^4 
+ 15 m_\pi^4 - 68 m_K^2 m_\pi^2 + (43 m_K^2 + 17 m_\pi^2) s_3 -24
\nonumber\\\et \times s_3^2) B_3^{\eta K}+\fr{1}{6}(673 m_K^4
- 794 m_K^2 m_\pi^2
+ 337 m_\pi^4 - \fr{36}{s_3}\Delta^2 (m_K^2 
 \nonumber\\\et + 8 m_\pi^2) - 6 (81 m_K^2+ 25 m_\pi^2) s_3 
+ 101 s_3^2) 
B_3^{ K\pi}-\fr{1}{2}(9 (5 m_K^4 
 \nonumber\\\et
- 6 m_K^2 m_\pi^2  + m_\pi^4) - 4 (m_K^2 
- 3 m_\pi^2) s_3
 - 5 s_3^2)
D_3^{\eta K}-\fr{1}{6}(477 m_K^4 
\nonumber\\\et + 369 m_\pi^4  + 1556 m_\pi^2 s_3 - 445 s_3^2
- m_K^2 (846 m_\pi^2 + 284 s_3))D_3^{ K\pi}
\Big] \,,\nonumber \\
\ea
\ba
S_6(s_1,s_3)&=&\fr{1}{216\Delta} \Big[
2 m_\pi^2 (17 m_K^2 - 5 m_\pi^2 - 9 s_1)B_1^{\eta\eta}
-18 \Big(4 m_K^4 + s_1 (2 s_3 + s_1) + 4
\nonumber\\\et
\times m_K^2  (3 m_\pi^2 - 2 (s_3 + s_1))\Big)B_1^{KK}
-9 (m_\pi^4 + 3 m_K^2 (9 m_\pi^2 - 7 s_1) + 3 s_3 s_1 
\nonumber\\\et
- m_\pi^2 (12 s_3 - 8 s_1))B_1^{\pi\pi}+\fr{1}{6}(332 m_K^4 - 177 m_\pi^4 + m_K^2 (781 m_\pi^2 - 594 s_3
\nonumber\\\et 
- 311 s_1)
 + 
  4 m_\pi^2 (27 s_3 - 13 s_1) + 3 (54 s_3 - 11 s_1) s_1) B_1^{\eta K}-\fr{1}{6}(9 \fr{\Delta^2}{s_1} (m_K^2
\nonumber\\\et 
- 7 m_\pi^2 - 6 s_3) + (335 m_K^4 + 104 m_\pi^4 - 27 m_\pi^2 s_3 - m_K^2 (439 m_\pi^2 - 81 s_3))
\nonumber\\\et
 \! -\! 3 (69 m_K^2 + 34 m_\pi^2 - 27 s_3) s_1 + 61 s_1^2) B_1^{ K\pi}\!+\! \fr{1}{2}(9
   \Delta (4 (m_K^2 + m_\pi^2) \!  
\nonumber\\\et\!-\! 3 s_3)
+ (7 m_K^2 + 39 m_\pi^2 + 27 s_3) s_1 -49 s_1^2)D_1^{\eta K}\! +\! \fr{1}{6}(9
\Delta (22 m_K^2 - 34 m_\pi^2 
\nonumber\\\et + 9 s_3)
 - (139 m_K^2 - 865 m_\pi^2 + 189 s_3) s_1 - 257 s_1^2  )D_1^{ K\pi}\Big] \, ,
\ea

\ba
B_n|_{Z_i=0} &=& 
\frac{C_{ew}}{f_{\pi}^2}\Big \lbrace 
\frac{2}{\Delta}\Big\lbrack L_3^r(-s_3(s_1+s_2)+2s_1s_2)
-2\,L_5^r(s_2+s_1-2s_3)m_{\pi}^2\Big\rbrack \nonumber\\
&&+\left( S_7(s_1,s_2,s_3)+S_8(s_1,s_3)+S_8(s_2,s_3)\right)
\Big \rbrace \,, {\rm with} \nonumber\\
S_7(s_1,s_2,s_3) &=& \frac{1}{144\Delta }\Big[\fr{6}{16\pi^2}( 4 m_K^4 + 14 m_K^2 m_\pi^2 +6 m_\pi^4- 6
s_3 (2 m_K^2+m_\pi^2)  
\nonumber\\ \et\! 
-3( 2 s_3^2
- s_1^2-  s_2^2))+
12 (s_1\!+s_2\!-2 s_3)\Big( 4 \nu_\eta\!+43 (\nu_K+\nu_\pi)\Big)
\nonumber\\ \et 
 -4\Big(\Delta (m_K^2-7 m_\pi^2)\Big(\fr{2}{s_3}- 
\fr{1}{s_1}-\fr{1}{s_2}\Big)
-\fr{12}{\Delta}   ( m_K^4+5 m_K^2m_\pi^2 \nonumber\\ \et\left.
+6 m_\pi^4-s_3 (4 m_K^2+5
m_\pi^2))\right)(\nu_K-\nu_\pi)
+\fr{2}{3 }\Big(18 (2 s_3-s_1-s_2 )
\nonumber\\ \et
 \times (s_3 B_3^{KK}\!-(2 m_\pi^2 
+ s_3)B_3^{\pi\pi})+(37 m_K^4+ 25 m_\pi^4- 134  m_K^2 m_\pi^2
\nonumber\\ \et
 + (54 m_\pi^2 + 90 m_K^2) s_3 - 63 s_3^2  ) 
 B_3^{\eta K}\!+6 ((-19 m_K^4 + 2 m_K^2 m_\pi^2 
\nonumber\\ \et
 + 29 m_\pi^4)+\frac{1}{s_3}
(m_K^6 - 9 m_K^4 m_\pi^2 + 15 m_K^2 m_\pi^4 - 7
m_\pi^6)+ 
 (25 m_K^2
\nonumber\\ \et
  - 19 m_\pi^2) s_3 - 3 s_3^2) B_3^{K\pi}+(135 m_K^4 + 6 m_K^2 (-27 m_\pi^2 + s_3) + 3 (9  
\nonumber\\ \et
\times m_\pi^2 - 5 s_3)(m_\pi^2 + 3 s_3))D_3^{\eta K}+3 (79 m_K^4 + 43
m_\pi^4 -122 m_\pi^2 m_K^2
\nonumber\\ \et
+ (110 m_\pi^2-86 m_K^2) s_3 + 15 s_3^2 ) D_3^{K\pi}\Big)\Big]\, ,
\ea
 
\ba
S_8(s_1,s_3) &=& \fr{1}{216\Delta}\Big[-\! 18 (2 m_K^4 + m_K^2 (6 m_\pi^2 - 4 s_3 - 3 s_1)
- s_1 (3 m_\pi^2 - s_3 - 2
 s_1))
\nonumber\\ \et\!
\times B_1^{KK}\!
- \!18 (m_K^2 (7 m_\pi^2 - s_1) + 21 m_\pi^4 - 12 m_\pi^2 (s_3 +s_1)
 + 3  s_3 s_1) B_1^{\pi\pi}
\nonumber\\ \et
+  (-50 m_K^4 + m_K^2 (-23 m_\pi^2 + 63
s_3) + (m_\pi^2 + 3 s_1)  (m_\pi^2 -\! 9 s_3 +6 s_1))
\nonumber\\ \et\!
\times B_1^{\eta K}\!  -3\Big(\! -19 m_K^4+ 20 m_\pi^4\! - m_K^2 (m_\pi^2 - 6
s_3)\!+\! \fr{1}{s_1}
  (m_K^6 - 9 m_K^4 m_\pi^2
\nonumber\\ \et
   + 15 m_K^2 m_\pi^4 
- 7 m_\pi^6)
 + 
  (16 m_K^2 + 5 m_\pi^2)   s_1- 18 s_1^2\Big) B_1^{K\pi}-3(\Delta (38
\nonumber\\ \et
\times  m_K^2- 26 m_\pi^2 + 3s_3)+  (-37 m_K^2 
+ 67 m_\pi^2 + 3 s_3) s_1 -21 s_1^2)D_1^{K\pi}
\nonumber\\ \et 
 -3 ( 3 \Delta (8 m_K^2-s_3)
+ (-5 m_K^2 - m_\pi^2 + 9 s_3) s_1
- 3 s_1^2)D_1^{\eta K}\Big] \, ,
\ea

\subsubsection{Electroweak Counterterm Contributions at $\order(e^2p^2)$}

The electroweak counterterm contributions at $\order(e^2p^2)$
break isospin symmetry and we use the decomposition in (\ref{AcBcdef}).
We define the constant
\be
C_{ew}^Z \,= \, - i\frac{3}{5}\frac{G_F}{\sqrt{2}}V_{ud}V_{us}^*
\fr{F_0^6}{f_K f_{\pi}^3}\re G_8 \, .
\ee

We get
\ba
A_{c,0}^Z(s_1,s_2,s_3)&=&  \frac{C_{ew}^Z}{f_\pi^2}\,
\left \lbrace -2 (m_K^2 + 5 m_\pi^2) Z^r_1- 2  (2 m_K^2 + 5 m_\pi^2) Z^r_2
+ \fr{4}{3}(m_K^2 - 2 m_\pi^2) Z^r_3 \right.\nonumber\\\et\hspace{1 cm}
+\fr{m_K^2}{9}\left(
-3Z^r_5
-2 Z^r_7+6 (Z^r_{11}+ Z^r_{12})\right)-(m_K^2 - 27 m_\pi^2) \fr{Z^r_6}{9}
 \nonumber\\\et \left.
\hspace{1 cm}- \fr{2 }{9}(m_K^2 - 3 m_\pi^2)
 (6 Z^r_{13}+ Z^r_{14})\right\rbrace\, , 
\ea
\ba
A_{c,+}(s_1,s_2,s_3)&=& A_{c,0}(s_1,s_2,s_3)-\frac{C_{ew}^Z}{f_\pi^2}
\fr{2 m_K^2}{3}
\left(-6 Z^r_{4} + 2 Z^r_{7} +  Z^r_{11}  +  Z^r_{12}\right)\, ,
\ea
\ba
B_c^Z(s_1,s_2,s_3)&=&  \frac{C_{ew}^Z}{f_\pi^2}
 \fr{(s_1+s_2-2 s_3)}{36\Delta} 
\Big \{ -18 (m_K^2 + m_\pi^2) Z^r_1 
- 18(2 m_K^2 + m_\pi^2)  Z^r_2
\nonumber\\\et
+\Delta \Big ( 
12(Z^r_3-Z^r_5)-31Z^r_6+4 Z^r_7+9 Z^r_8-48 Z^r_{13}-8Z^r_{14}\Big )
\nonumber\\\et
- 9 (3 m_K^2 - 2 m_\pi^2) Z^r_9 -6 m_K^2 Z^r_{10}+ 6 (4 m_K^2 - 3 m_\pi^2)
Z^r_{11}
\nonumber\\\et 
-6 (4 m_K^2 - 5 m_\pi^2) Z^r_{12}\Big\}\, ,
\ea
\ba
B_{t,0}^Z(s_1,s_2,s_3)&=&\frac{C_{ew}^Z}{f_\pi^2}
 \fr{s_1+s_2-2s_3}{12\Delta}\Big\{
6 (m_K^2 + m_\pi^2)Z^r_1+6 (2 m_K^2 + m_\pi^2) Z^r_2
\nonumber\\\et -\Delta ( 4  Z^r_3\! +  3  Z^r_6)
+ 3 (3 m_K^2  - 2 m_\pi^2) Z^r_8 
- 3 (m_K^2 - 2 m_\pi^2) Z^r_9\nonumber\\\et + 2 m_K^2 Z^r_{10} 
- 2  m_\pi^2 Z^r_{11}
+2 (4 m_K^2  - 5 m_\pi^2) Z^r_{12} \Big\}\, ,
\ea
\ba
B_{t,+}^Z(s_1,s_2,s_3)&=& B_{t,0}^Z(s_1,s_2,s_3)
+\frac{C_{ew}^Z}{f_\pi^2}\fr{s_1+s_2-2 s_3}{3}(2
Z_{11}^r-4Z_{12}^r+Z_7^r)\, ,
\ea
\ba
A_{n,0}^Z(s_1,s_2,s_3) &=& \frac{C_{ew}^Z}{f_\pi^2}
\,\fr{m_K^2}{9}\left(  3 Z^r_{5} 
- 2 Z^r_{6} + 2 Z^r_{7} - 3 Z^r_{8} - 
3 Z^r_{9}-2 Z^r_{10}\right.\nonumber\\\et
\hspace{1 cm}\left.+2  Z^r_{11} + 2 Z^r_{12}\right)\,,
\ea
\ba
A_{n,+}^Z(s_1,s_2,s_3) &=&  A_{n,0}^Z(s_1,s_2,s_3)
+\frac{C_{ew}^Z}{f_\pi^2}\fr{2 m_K^2}{3}\left(-6 Z^r_{4} + 2 Z^r_{7} 
+  Z^r_{11}  +  Z^r_{12}\right)\, ,
\ea
\ba
B_{n}^Z(s_1,s_2,s_3) &=& -\frac{C_{ew}^Z}{f_\pi^2}
\fr{s_1+s_2-2s_3}{18\Delta}\left\{
18 (m_K^2+m_\pi^2) Z^r_1+18 (2 m_K^2+m_\pi^2) Z^r_2\right.
\nonumber\\\et
-\Delta (12  Z^r_3+6 Z^r_5+5
 Z^r_6-8 Z^r_7)-3 ( m_K^2-4 m_\pi^2) (Z^r_8 +Z^r_9)
\nonumber\\\et\left.
-2 (m_K^2-4
 m_\pi^2)Z^r_{10}
+2 ( 4 m_K^2-7 m_\pi^2) (Z^r_{11}+Z^r_{12})\right\}\, .
\ea

\section{The Slope $g$ and $\Delta g$ at LO and NLO} 
\label{Adeltag}

We have checked that the following relations
\begin{itemize}
  \item $F_0^2\, \re \left(e^2G_E\right)<<m_\pi^2\re G_8$ 
  \item $\im G_8<<\re G_8$
  \item $\im (e^2 G_E)<<\re G_8$
\end{itemize}
can be used in this and the next sections to simplify the analytical 
expressions. To obtain the numerical results included in the
 text we use the full expressions, with no simplifications. We have also
checked that the terms 
disregarded with the application of these relations generate very small 
changes in the numbers.

Using the simplifications above, 
the value of $g$ at LO can be written trivially as 
\be \label{gLO}
g^{\rm LO}\,=\,2\frac{B_{8,1}^{(2)}\re G_8+B_{27,1}^{(2)}G_{27}
+B_{E,1}^{(2)}\re \left(e^2G_E\right)}
{B_{8,0}^{(2)}\re G_8 +B_{27,0}^{(2)}G_{27}
+B_{E,0}^{(2)}\re \left(e^2G_E\right)}\,.
\ee
The expressions for $B_{8,j}^{(2)}$, $B_{27,i}^{(2)}$,
and $B_{E,i}^{(2)}$ needed above can be obtained from
the expressions of the corresponding $B$'s for the
charged Kaon decays $++-$ and $00+$  in Appendix \ref{ANLO}
 and expanding them 
as in (\ref{yexpansion}). The results we  get 
are in (\ref{gLOCN}). 

We consider now the NLO corrections to the slope  $g$. 
Disregarding the tiny CP-violating we have $g[K^+\to 3\pi]=
g[K^-\to 3\pi]$ at NLO we get 
\ba \label{AgNLO}
A_0^{\rm NLO} &=& \left\lbrack \sum _{i=8,27}\left(B_{i,0}^{(2)}
+B_{i,0}^{(4)}\right)\re G_i 
\,+\,\sum_{j=1,11}H_{j,0}^{(4)}\re \widetilde K_j
\right\rbrack^2
\,+\,\left\lbrack \sum _{i=8,27}\left(C_{i,0}^{(4)}\re G_i\right)
\right\rbrack^2\, ,\nonumber\\
A_y^{\rm NLO} &=& 2\,\Bigg\lbrace
 \left\lbrack \sum _{i=8,27}\left(B_{i,0}^{(2)}
+B_{i,0}^{(4)}\right)\re G_i \,+\,\sum_{j=1,11}H_{j,0}^{(4)}\re \widetilde K_j
\right\rbrack\nonumber\\ &&\times
\left\lbrack \sum _{i=8,27}\left(B_{i,1}^{(2)}
+B_{i,1}^{(4)}\right)\re G_i \,+\,\sum_{j=1,11}H_{j,1}^{(4)}\re \widetilde K_j
\right\rbrack
\nonumber\\
&&+\left\lbrack \sum _{i=8,27}\left(C_{i,0}^{(4)}\re G_i\right)
\right\rbrack
\times\left\lbrack \sum _{i=8,27}\left(C_{i,1}^{(4)}\re G_i\right)
\right\rbrack
\Bigg \rbrace\,
\ea
for the coefficients defined in (\ref{amp2}).

One can get   $g_{C(N)}$ at NLO 
using (\ref{AmpDeltag}) and the results above
  substituting the coefficients $C_{i,j}^{(4)}$, $B_{i,j}^{(2n)}$ and 
$H_{i,j}^{(4)}$ by their values calculated expanding in the Dalitz variables 
the results in \cite{BDP03}.
 
 The slope $g$ asymmetry  in (\ref{defDeltag}) can be written at LO as
\be 
\Delta g_{C(N)} \simeq \frac{m_K^2}{F_0^2} \, B_{C(N)} \, \im G_8 +
D_{C(N)} \, \im (e^2 G_E) \, .
\ee
We get
\ba \label{ACBC}
B_C &=& -\frac{15}{64}\frac{1}{\pi}\,G_{27}
\sqrt{\frac{\mkd-9\mpd}
{\mkd+3\mpd}}\,\nonumber\\
&&\times \frac{14\mkd\mpd-18m_{\pi}^4+5m_K^4}{\mkd(\mkd-\mpd)
(3\re G_8+2G_{27})(13G_{27}-3\re G_8 )}\, ,\nonumber\\
D_C &=& \frac{3}{64}\frac{1}{\pi}\,\sqrt{\frac{\mkd-9\mpd}
{\mkd+3\mpd}}\frac{1}{\mkd(\mkd-\mpd)(3\re G_8 +2G_{27})
(13G_{27}-3\re G_8 )}
\nonumber\\ &&
\times  \left\lbrack3\re G_8 (16\mkd\mpd-18m_{\pi}^4+3m_K^4)
-G_{27}(178\mkd\mpd-234m_{\pi}^4+69m_K^4)\right\rbrack\,,\nonumber\\
\ea
and, in the neutral case,
\ba \label{ANBN}
B_N &=&  -\frac{15}{32}\frac{1}{\pi}\,G_{27}
\sqrt{\frac{\mkd-9\mpd}{\mkd+3\mpd}}\frac{7\mpd+4\mkd}{E}
\,,\nonumber\\
D_N &=& \frac{9}{32}\frac{1}{\pi}\,\re G_{8} 
\sqrt{\frac{\mkd-9\mpd}{\mkd+3\mpd}}\frac{\mpd(18\mpd-7\mkd)}{\mkd E}
\nonumber\\
&&+\frac{3}{32}\frac{1}{\pi}\,G_{27}
\sqrt{\frac{\mkd-9\mpd}{\mkd+3\mpd}}\frac{36m_{\pi}^4-119\mpd\mkd
-60m_K^4}{\mkd E}\,,
\ea
with 
\be 
E  \equiv (3\re G_8 +2G_{27})\left((19\mkd-4\mpd)G_{27}+6(\mkd-\mpd)
\re G_8 \right)\,.
\ee
The sum $A^+_yA^-_0 + A^+_0 A^-_y$, necessary to get
$\Delta g$ at NLO using (\ref{amp2}) and (\ref{AmpDeltag}), can 
be obtained directly from
(\ref{AgNLO}) where we have neglected the small CP-violating
effects. 

For the difference  $A^+_yA^-_0 - A^+_0 A^-_y$, we get 
\ba 
\left(A^+_yA^-_0 - A^+_0 A^-_y \right)_{\rm NLO}
&=& 4 {\cal A}_I\left\lbrack
\left({\cal A}_R^2-{\cal C}_R^2\right){\cal D}_R-2{\cal A}_R
{\cal B}_R{\cal C}_R
\right\rbrack+4 {\cal C}_I\left\lbrack
\left({\cal A}_R^2-{\cal C}_R^2\right){\cal B}_R
\right. \nonumber\\  &&\left.
+2{\cal A}_R
{\cal C}_R{\cal D}_R
\right\rbrack+4\left({\cal B}_I{\cal C}_R-{\cal D}_I{\cal A}_R\right)
\left({\cal A}_R^2+{\cal C}_R^2\right)
\, , 
\ea
where ${\cal A}_R$, ${\cal B}_R$, ${\cal C}_R$ and ${\cal D}_R$ 
contain the contributions from the real parts 
of the counterterms 
\ba
{\cal A}_R &=& \sum_{i=8,27,E} 
\left(B_{i,0}^{(2)}+B_{i,0}^{(4)}\right)\, \re G_i\, 
+\, \sum_{i=1,11}H_{i,0}^{(4)} \, \re \widetilde K_i \, ,\nonumber\\
{\cal B}_R &=& \sum_{i=8,27,E} 
\left(B_{i,1}^{(2)}+ B_{i,1}^{(4)}\right)\,
\re G_i\, +\,
  \sum_{i=1,11}H_{i,1}^{(4)}\, \re \widetilde K_i \,  ,\nonumber\\
{\cal C}_R &=& \sum_{i=8,27,E} 
\left(C_{i,0}^{(4)} +C_{i,0}^{(6)}\right)  \,
\re G_i\, 
+\,\sum_{i=1,11}H_{i,0}^{(6)} \, \re \widetilde K_i\, ,\nonumber\\
{\cal D}_R &=& \sum_{i=8,27,E} 
\left( C_{i,1}^{(4)} + C_{i,1}^{(6)}\right)  \,
\re G_i\, +\,\sum_{i=1,11}H_{i,1}^{(6)}\, \re \widetilde K_i\, .
\ea
While  ${\cal A}_I$, ${\cal B}_I$, ${\cal C}_I$ 
are the same expressions but substituting 
the real parts of the counterterms by their imaginary parts.

The coefficients  $B_{i,0(1)}^{(2n)}$, $C_{i,0(1)}^{(2n)}$ and 
$H_{i,0(1)}^{(2n)}$  defined in (\ref{yexpansion}) are real.

\section{The Quantities  $|A|^2$ and $\Delta |A|^2$ at LO and NLO} 
\label{ANLO}

Here we give the results for the quantities $A$ and $\Delta A$
defined in (\ref{defACN}) and (\ref{Deltapm}), respectively. 
They enter in the integrands of the decay rates 
$\Gamma$  in (\ref{eq:extrs})  and the CP-violating asymmetries
$\Delta \Gamma$, see (\ref{eq:dGLO}).

To simplify the analytical expressions, 
we have made use
 of the fact that  the imaginary  part of the counterterms 
is much smaller than their  real parts. 
The $|A_{C(N)}|^2$  which give the asymmetries 
$\Delta \Gamma$ at LO are in  (\ref{GammaLO}).  

The result for $\Delta |A_{C}|^2$ at LO
can be obtained  substituting  in (\ref{DGammaLO}) 
the functions $B_i^{(2)}(s_1,s_2,s_3)$ and $C_i^{(4)}(s_1,s_2,s_3)$  
for $i=8$ and $i=27$  by
\ba \label{DeltapmLOppm}
B_{8\,++-}^{(2)}(s_1,s_2,s_3) &=& i\,\frac{C\,F_0^4}{f_{\pi}^3f_K}
\,(s_3-\mkd-\mpd)\, ,\nonumber\\
B_{27\,++-}^{(2)}(s_1,s_2,s_3) &=& i\,\frac{C\,F_0^4}{f_{\pi}^3f_K}
\,\frac{1}{3}(13\mpd+3\mkd-13s_3)\, ,\nonumber\\
C_{8\,++-}^{(4)}(s_1,s_2,s_3) &=&  i\,\frac{C\,F_0^4}{f_{\pi}^3f_K}\,
\left(-\frac{1}{16\pi f_\pi^2} \right)\nonumber\\
&&\times\Bigg \lbrace \frac{1}{2}
\left\lbrack s_3^2-s_3(3m_{\pi}^2+m_K^2)+2m_{\pi}^2(m_K^2+m_{\pi}^2)
\right\rbrack\sigma(s_3)\nonumber\\
&+&
\frac{1}{6}\left\lbrack 4s_2^2+s_2(-4m_{\pi}^2+2m_K^2-s_3) 
+m_{\pi}^2(4s_3-2m_K^2-3m_{\pi}^2)\right\rbrack
\sigma(s_2)\nonumber\\
&+&({\rm exchange\,\, s_1 \,\,  and\,\, s_2\,\,  in\,\, the\,\, 
 second\,\, term})\Bigg\rbrace\,,\nonumber\\
C_{27\,++-}^{(4)}(s_1,s_2,s_3) &= &i\,\frac{C\,F_0^4}{f_{\pi}^3f_K}\,
\left(-\frac{1}{16\pi f_\pi^2} \right)\nonumber\\
&&\times\Bigg \lbrace \frac{1}{6}\left\lbrack-13s_3^2+3s_3(13m_{\pi}^2
+m_K^2)-2m_{\pi}^2(3m_K^2+13m_{\pi}^2)\right\rbrack
\sigma(s_3)\nonumber\\
&&+\frac{1}{36(m_K^2-m_{\pi}^2)} \Big\lbrack s_2^2(14m_{\pi}^2+31m_K^2)
+s_2(26s_3(m_K^2-m_{\pi}^2)\nonumber\\
&&-174m_K^2m_{\pi}^2-7m_K^4+76m_{\pi}^4)+(104m_{\pi}^2s_3(m_{\pi}^2-m_K^2)
\nonumber\\
&&-168m_{\pi}^6+161m_K^2m_{\pi}^4+67m_K^4m_{\pi}^2) \Big\rbrack \,
\sigma(s_2)\nonumber\\
&+&({\rm exchange\,\, s_1 \,\,  and\,\, s_2\,\,  in\,\, the\,\, 
 second\,\, term}) 
\Bigg \rbrace \,,
\ea
and $B_E^{(2)}$, $C_E^{(4)}$ by
\ba \label{DeltapmLOppmEM}
B_{E\,++-}^{(2)} &=& i\,\frac{C\,F_0^4}{f_{\pi}^3f_K}
\, \left(-2F_0^2\right)\,,\nonumber\\
C_{E\,++-}^{(4)} &=& i\,\frac{C\,F_0^4}{f_{\pi}^3f_K}\, 
\left(\frac{-F_0^2}{16\pi f_\pi^2}\right) 
\,\Bigg \lbrace (2m_{\pi}^2-s_3)\sigma(s_3)
\nonumber\\
&&+\frac{1}{4(m_K^2-m_{\pi}^2)} \left \lbrack 3s_2^2+s_2
(5m_K^2-12m_{\pi}^2)+m_{\pi}^2(5m_{\pi}^2-m_K^2)\right\rbrack
\sigma(s_2)\nonumber\\
&&+ ({\rm exchange\,\, s_1 \,\,  and\,\, s_2\,\,  in\,\, the\,\, 
 second\,\, term}) \Bigg \rbrace \, . 
\ea

One can get $\Delta |A_N|^2$ at LO substituting  in (\ref{DGammaLO}) 
the functions $B_i^{(2)}(s_1,s_2,s_3)$ and $C_i^{(4)}(s_1,s_2,s_3)$  
for $i=8$  and $i=27$ by  
\ba\label{DeltapmLO00p}
B_{8\,00+}^{(2)}(s_1,s_2,s_3) &=& i\,\frac{C\,F_0^4}{f_{\pi}^3f_K}
\,(\mpd-s_3)\, ,\nonumber\\
B_{27\,00+}^{(2)}(s_1,s_2,s_3) &=& i\,\frac{C\,F_0^4}{f_{\pi}^3f_K}
\,\frac{1}{6(\mkd-\mpd)}\left\lbrack 5m_K^4+19\mpd\mkd-4m_{\pi}^4
+s_3(4\mpd-19\mkd)\right\rbrack\, ,\nonumber\\
C_{8\,00+}^{(4)}(s_1,s_2,s_3) &=&  i\,\frac{C\,F_0^4}{f_{\pi}^3f_K}\,
\left(-\frac{1}{16\pi f_\pi^2} \right)\nonumber\\
&&\times\Bigg \lbrace \frac{1}{2}
\left\lbrack s_3^2+s_3(m_K^2-m_{\pi}^2)-m_{\pi}^2m_K^2
\right\rbrack\sigma(s_3)\nonumber\\
&+&
\frac{1}{6}\left\lbrack 2s_2^2+s_2(s_3-2(4m_{\pi}^2+m_K^2)) 
+m_{\pi}^2(-4s_3+5m_K^2+9m_{\pi}^2)\right\rbrack
\sigma(s_2)\nonumber\\
&+& ({\rm exchange\,\, s_1 \,\,  and\,\, s_2\,\,  in\,\, the\,\, 
 second\,\, term}) \Bigg\rbrace\,,\nonumber\\
C_{27\,00+}^{(4)}(s_1,s_2,s_3) &= &i\,\frac{C\,F_0^4}{f_{\pi}^3f_K}\,
\left(-\frac{1}{16\pi f_\pi^2} \right)\frac{1}{(\mkd-\mpd)}\nonumber\\
&&\times\Bigg \lbrace \frac{-1}{12}\Big\lbrack 26s_3^2(m_K^2-m_{\pi}^2)
+s_3(56m_{\pi}^4-57m_K^2m_{\pi}^2-14m_K^4)\nonumber\\
&&\hspace{0.5cm}+m_{\pi}^2(31m_K^2m_{\pi}^2-30m_{\pi}^4+19m_K^4)\Big\rbrack
\sigma(s_3)\nonumber\\
&&+\frac{1}{36} \Big\lbrack s_2^2(-8m_{\pi}^2+38m_K^2)
+s_2(s_3(19m_K^2-4m_{\pi}^2)\nonumber\\
&&\hspace{0.5 cm} -144m_K^2m_{\pi}^2-23m_K^4+32m_{\pi}^4)
+s_3(16m_{\pi}^2-76m_K^2)m_{\pi}^2\nonumber\\
&&\hspace{0.5 cm}-36m_{\pi}^6+151m_K^2m_{\pi}^4+65m_K^4m_{\pi}^2) \Big\rbrack 
 \sigma(s_2)\nonumber\\
&&+ ({\rm exchange\,\,  s_1\,\,   and \,\, s_2 \,\, in\,\, the \,\, 
second\,\, term}) \Bigg \rbrace \,,
\ea
and 
\ba \label{DeltapmLO00pEM}
B_{E\,00+}^{(2)} &=& i\,\frac{C\,F_0^4}{f_{\pi}^3f_K}
\, \frac{F_0^2}{2(\mkd-\mpd)}\left(5\mpd-\mkd-3s_3\right)\,,\nonumber\\
 C_{E\,00+}^{(4)} &=& i\,\frac{C\,F_0^4}{f_{\pi}^3f_K}\, 
\left(\frac{-1}{16\pi(\mkd-\mpd)} \right)\nonumber\\
&&\times\Bigg \lbrace \frac{1}{4}\left\lbrack s_3(8m_K^2-5m_{\pi}^2)
+m_{\pi}^2(3m_{\pi}^2-7m_K^2)\right\rbrack\sigma(s_3)
\nonumber\\
&&+\frac{1}{4} \left \lbrack 2s_2^2+s_2
(s_3-3(m_K^2+2m_{\pi}^2))+m_{\pi}^2(-4s_3+5m_{\pi}^2+7m_K^2)\right\rbrack
\sigma(s_2)\nonumber\\
&&+\frac{1}{4} \left \lbrack 2s_1^2+s_1
(s_3-3(m_K^2+2m_{\pi}^2))+m_{\pi}^2(-4s_3+5m_{\pi}^2+7m_K^2\right\rbrack
\sigma(s_1) \Bigg \rbrace\,. \, 
\ea

The function $\sigma(s)$ appearing in all the formulas
above is
\be
\label{sigma}
\sigma(s)\,=\,\sqrt{1-\frac{4\mpd}{s}}\, .
\ee
In all the expressions at LO
we  use $f_K=f_\pi=F_0$.

At NLO,  we get
\ba
\label{eqANLO}
|A_{NLO}|^2 &=&  
\left\lbrack\left(B_8^{(2)}+B_8^{(4)}\right)\re G_8
+\left(B_{27}^{(2)}+B_{27}^{(4)}\right)G_{27}+\sum_{i=1,11}H_i^{(4)}\re 
\widetilde K_i
\right\rbrack^2\nonumber\\
&&+\left\lbrack C_8^{(4)}\re G_8 +C_{27}^{(4)}G_{27}\right\rbrack^2
\, ,\nonumber\\
\Delta |A_{NLO}|^2 &=&  
-2 \im G_8 \Bigg \lbrace \left\lbrack G_{27}\left(B_{27}^{(2)}+B_{27}^{(4)}
\right)+\sum_{i=1,11}H_i^{(4)}\re 
\widetilde K_i\right\rbrack\left(C_8^{(4)}+C_8^{(6)}\right)
\nonumber\\
&&\hspace{1. cm}-\left(B_{8}^{(2)}+B_{8}^{(4)}
\right)\left\lbrack G_{27}\left(C_{27}^{(4)}+C_{27}^{(6)}\right)
+H_i^{(6)}\re \widetilde K_i\right\rbrack
\Bigg\rbrace\nonumber\\
&&-2\im \left(e^2G_E\right)\Bigg \lbrace \left( \sum_{i=8,27}\left\lbrack
\re G_i\left(B_i^{(2)}+B_i^{(4)}
\right)\right\rbrack+\sum_{i=1,11}H_i^{(4)}\re 
\widetilde K_i\right)\,C_E^{(4)}\nonumber\\
&&\hspace{1. cm}-\left(B_{E}^{(2)}+B_{E}^{(4)}\right)\left( 
\sum_{i=8,27}\left\lbrack\re G_{i}\left(C_{i}^{(4)}+C_{i}^{(6)}\right)
\right\rbrack+\sum_{i=1,11}H_i^{(6)}\re \widetilde K_i\right)
\Bigg\rbrace\nonumber\\
&&+\left(2\sum_{i=1,11}H_i^{(4)}\im 
\widetilde K_i \right) \Bigg\lbrace
\sum_{i=8,27}\left\lbrack\re G_i \,\left(C_i^{(4)}+C_i^{(6)}
\right)\right\rbrack \nonumber\\
&&\hspace{7 cm}
+\sum_{i=1,11}H_i^{(6)}\re \widetilde K_i\Bigg\rbrace\nonumber\\
&&-2\left(\sum_{i=1,11}H_i^{(6)}\im \widetilde K_i\right)\Bigg\lbrace
\sum_{i=8,27}\left\lbrack\re G_i \,\left(B_i^{(2)}+B_i^{(4)}
\right)\right\rbrack \nonumber\\
&&\hspace{7 cm}+\sum_{i=1,11}H_i^{(4)}\re \widetilde K_i\Bigg\rbrace
\,.
\ea
Again, we disregarded the $s_i$ dependence of the functions $B_i^{(2n)}$, 
$C_i^{(2n)}$ and $H_i^{(2n)}$. The functions $B_{8(27)}^{(4)}$ and 
$H_i^{(4)}$ can be deduced from the results in \cite{BDP03} and the functions 
$B_{E}^{(4)}$ from Subsection \ref{ZiK3pi}. Finally, the functions 
$C_i^{(6)}$ and $H_i^{(6)}$ are discussed in Appendix \ref{FSI6}.

\section{Final State Interactions at NLO} 
\label{FSI6}

In this Appendix we provide some details of the calculation 
of the FSI using
the optical theorem in the framework of CHPT. We compute the imaginary
part of the amplitudes at ${\cal O}(p^6)$. 
The calculation corresponds to the diagrams shown 
in Figures \ref{fig:imp6ppm} and \ref{fig:imp6p00}.
\FIGURE{\epsfig{file=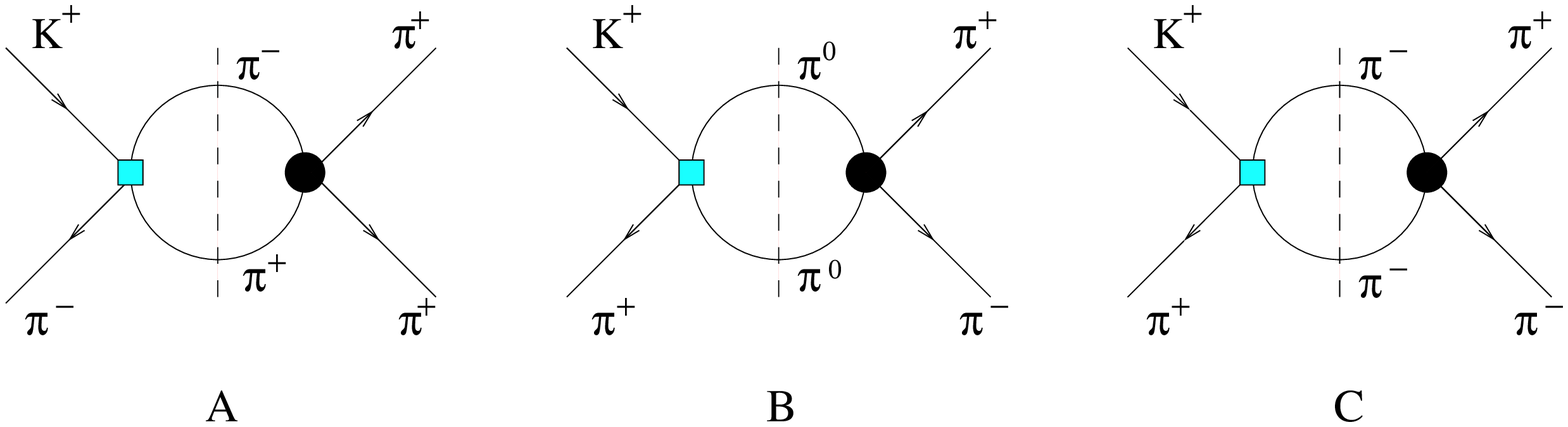,width=10cm}
        \caption{Relevant diagrams for the calculation of FSI for
 $K^+\rightarrow\pi^+\pi^+\pi^-$. 
The square vertex is the weak vertex and the
 round one is  the strong vertex}
\label{fig:imp6ppm}}
\FIGURE{\epsfig{file=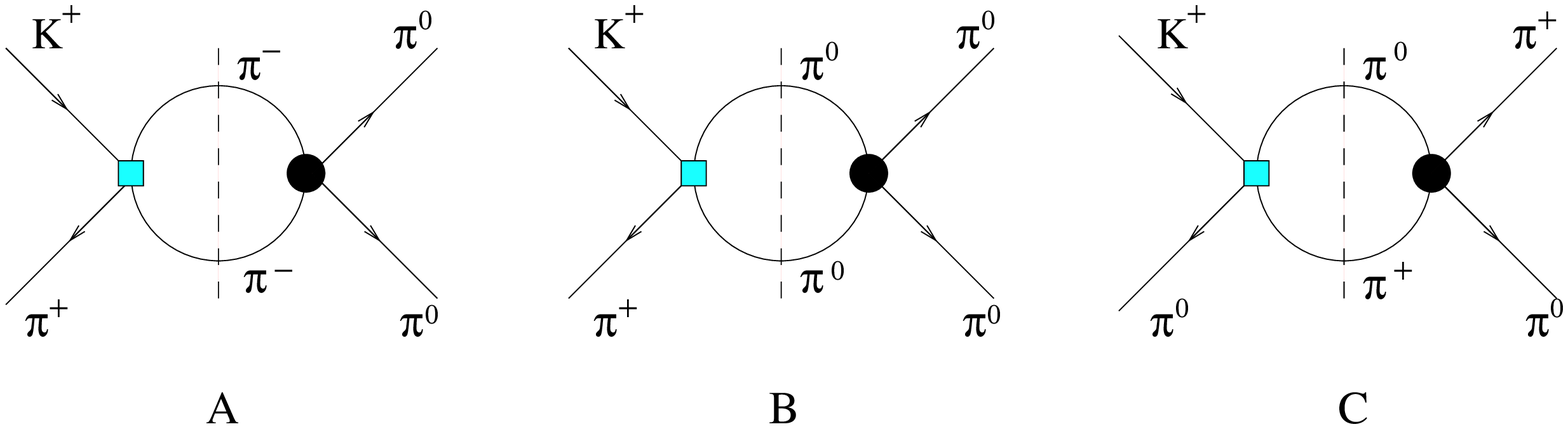,width=10cm}
        \caption{Relevant diagrams for the calculation of FSI for
 $K^+\rightarrow\pi^0\pi^0\pi^+$. The square vertex is the weak 
vertex and the
 round one is  the strong vertex.}
\label{fig:imp6p00}}
We can distinguish the cases in which the weak vertex 
is of $ {\cal O}(p^4)$
and the strong vertex of order  $ {\cal O}(p^2)$ and the inverse 
case in which
the weak vertex is of order  $ {\cal O}(p^2)$ and the strong 
vertex of order 
 $ {\cal O}(p^4)$. In this paper we will not consider the weak 
vertices  generated by the electroweak penguin.
In  Subsection \ref{sec:not6} we provide  some notation. In Subsections 
\ref{sec:ppm6} and \ref{sec:p006}
we report the  calculation for the  charged Kaon decays.
An example of  the calculation of the integrals 
that must be performed is given in Subsection 
\ref{sec:integ6}. Finally, in 
Subsection \ref{phasesNLO} we give analytical results for 
the strong phases at NLO.

\subsection{Notation}
\label{sec:not6}
In order to be concise we use  the functions $M_i$  for the weak 
amplitudes given in \cite{BDP03}. We define
\eq\label{asbs}
\widetilde{M}_i(s)&=& \int_{-1}^1 d\cos\theta\;  \left. M_i(a(s)+b(s) \cos\theta)\right|_{p^4}\, ,\\
\widetilde{M}_i^s(s)&=& \int_{-1}^1 d\cos\theta\;  \left.  (a(s)+b(s) \cos\theta)
M_i(a(s)+b(s) \cos\theta)\right|_{p^4}\,,\\
\widetilde{M}_i^{ss}(s)&=& \int_{-1}^1 d\cos\theta\; \left.  (a(s)+b(s) \cos\theta)^2 M_i(a(s)+b(s) \cos\theta)\right|_{p^4}\,,\\
a(s)&=& \fr{1}{2}( m_K^2+3m_\pi^2-s)\,,\nn\\
b(s)&=& \fr{1}{2}\sqrt{(s-4 m_\pi^2)\left(s-2  (m_K^2+m_\pi^2)+\fr{(m_K^2-m_\pi^2)^2}{s}\right)}\,.
\en

The amplitudes at ${\cal O}(p^4)$ for the $\pi\pi\rightarrow \pi\pi$ 
scattering  in a theory with three flavors can be found  in \cite{BKM91}.
We decompose the amplitudes in the various cases as follows.
For the case $\pi^+\pi^+\rightarrow \pi^+\pi^+$ the amplitude at ${\cal O}(p^4)$
is
\eq
\Pi_1&=& P_1(s)+P_2(s,t)+P_2(s,u) .
\en
For the case $\pi^0\pi^0\rightarrow \pi^+\pi^-$ the amplitude at ${\cal O}(p^4)$
is
\eq
\Pi_2&=& P_3(s)+P_4(s,t)+P_4(s,u) .
\en
For the case $\pi^+\pi^-\rightarrow \pi^+\pi^-$ the amplitude at ${\cal O}(p^4)$
is
\eq
\Pi_3&=& P_5(s)+P_6(s,t)+P_6(s,u)+P_7(s,t)-P_7(s,u) .
\en
Finally the amplitude
$\pi^0\pi^0\rightarrow \pi^0\pi^0$ at ${\cal O}(p^4)$
is
\eq
\Pi_4&=& P_8(s)+P_8(t)+P_8(u) .
\en
The value for the various $P_i$ can be deduced from \cite{BKM91}.
In the following we use
\eq\!\!\!\!\!\!\!\!\!
\widetilde{P}^{(n,m)}_i(s)&=& \int_{-1}^1 d\cos\theta\;s^n  
(c(s) (1- \cos\theta))^m 
P_i(s,c(s) (1- \cos\theta))\,,\\
\!\!\!\!\!\!\!\!\!
\widehat{P}^{(n)}_{1,i}(s)&=&\int_{-1}^1 d\cos\theta\;(c(s)(1-\cos\theta))^n
P_i(c(s)(1+\cos\theta),c(s)(1-\cos\theta))\,, \\
\!\!\!\!\!\!\!\!\!
\widehat{P}^{(n)}_{2,i}(s)&=&\int_{-1}^1 d\cos\theta\;(c(s)(1-\cos\theta))^n
P_i(c(s)(1-\cos\theta),s)\,, \\
\!\!\!\!\!\!\!\!\!
c(s)&=&-\fr{1}{2} (s-4 m_\pi^2)\,.
\en
Another function we use in the next subsections  is
$\sigma(s)$ which was defined already in (\ref{sigma}).

\subsection{Final State Interactions for $K^+\rightarrow\pi^+\pi^+\pi^-$}
\label{sec:ppm6}

We first  compute the  contributions
depicted in Figure  \ref{fig:imp6ppm} in which the weak vertex is of
${\cal O}(p^4) $ and the strong vertex of ${\cal O}(p^2) $.
 The results for the diagrams A and B are
\eq
{\rm Im}A^{(6,1)}_W&=& \fr{\s(s_3)}{32\pi}
\frac{(2 m_\pi^2-s_3)}{f_{\pi}^2}\left[
  \left. M_{10}(s_3)\right|_{p^4}+ \widetilde{M}_{11}(s_3)+ 
\widetilde{M}_{12}(s_3) (m_K^2\right. \nonumber \\
&&+3 m_\pi^2-2
  s_3)\left. - \widetilde{M}_{12}^s(s_3)\right]
\,,\\
{\rm Im}A^{(6,2)}_W&=& \fr{\s(s_1)}{32\pi}\frac{( s_1-m_\pi^2)}{f_{\pi}^2}\left[
  \left.M_{7}(s_1)\right|_{p^4}+\widetilde{M}_{8}(s_1)+\widetilde{M}_{9}(s_1) (m_K^2+3 m_\pi^2-2
  s_1)\right.\nonumber \\
&&
\left. -\widetilde{M}_{9}^s(s_1)\right] 
 +(s_1\leftrightarrow s_2)
\,,
\en
respectively. For diagram C we have both $S$-wave  and 
$P$-wave contributions. We get for them  
\eq 
\label{eq:im6,3ws}
{\rm Im}A^{(6,3)}_{W,S}&=& \fr{\s(s_1)}{64\pi} \frac{s_1}{f_{\pi}^2}\left[
  2
  \left. M_{11}(s_1)\right|_{p^4}+\widetilde{M}_{11}(s_1)+\widetilde{M}_{10}(s_1)+\widetilde{M}_{12}^s(s_1) \right. \nonumber \\ &&
\left.
-\widetilde{M}_{12}(s_1) (m_K^2+3 m_\pi^2-2
  s_1) \right]  +(s_1\leftrightarrow s_2)
\,,  \\ 
\label{eq:im6,3wp}
{\rm Im}A^{(6,3)}_{W,P}&=& \fr{\s(s_1)}{64\pi}\frac{1}{f_{\pi}^2} \fr{s_1 (s_3-s_2)}{s_1^2-2 (m_K^2+m_\pi^2) s_1+(m_K^2-m_\pi^2)^2}
\left[(s_1-(m_K^2+3 m_\pi^2))\right. \nonumber \\
\et
\times(\widetilde{M}_{11}(s_1)-\widetilde{M}_{10}(s_1)+\widetilde{M}_{12}(s_1)(2
s_1-m_K^2-3  m_\pi^2)) 
+2 \widetilde{M}^s_{11}(s_1) \nonumber \\
\et-2 \widetilde{M}^s_{10}(s_1)+\widetilde{M}^s_{12}(s_1)(5 s_1-3(m_K^2+3
  m_\pi^2))+2 \widetilde{M}^{ss}_{12}(s_1) \nonumber \\
\et
+\fr{8}{3} b^2(s_1) \left. M_{12}(s_1)\right|_{p^4}
  \Big] 
 +(s_1\leftrightarrow s_2)\, , 
\en
respectively.

Secondly, 
 we report the calculation of  the  case in which the strong vertex
 is of ${\cal O}(p^4)$ and the weak vertex is of ${\cal O}(p^2)$.
With analogous notation as above, we get
\eq
{\rm Im}A^{(6,1)}_\pi&=& \fr{\s(s_3)}{32\pi} 
 \left.M_{10}(s_3)\right|_{p^2}\ 
(P_1(s_3)+\widetilde{P}^{(0,0)}_2(s_3))\,,
\\ 
{\rm Im}A^{(6,2)}_\pi&=& \fr{\s(s_1)}{32\pi} \left. (M_{7}(s_1)+M_{8}(s_2)+M_{8}(s_3))\right|_{p^2}\, 
(P_3(s_1)+\widetilde{P}^{(0,0)}_4(s_1))\nonumber
\\ \et
+(s_1\leftrightarrow s_2) \,,
\\ 
{\rm Im}A^{(6,3)}_{\pi,S}&=& \fr{\s(s_1)}{32\pi}\left(\left.M_{10}(s_3)\right|_{p^2}+\left.M_{10}(s_2)\right|_{p^2}\right)
(P_5(s_1)+\widetilde{P}^{(0,0)}_6(s_1))
\nonumber \\ \et
+(s_{1}\leftrightarrow s_{2}) \,,
\\
{\rm Im}A^{(6,3)}_{\pi,P}&=&
\fr{\s(s_1)}{32\pi}\left(\left.M_{10}(s_3)\right|_{p^2}-\left.M_{10}(s_2)\right|_{p^2}\right)\fr{1}{s_1-4 m_\pi^2}
\Big( (s_1-4 m_\pi^2)\widetilde{P}_7^{(0,0)}(s_1)
\nonumber \\ \et
+2\widetilde{P}_7^{(0,1)}(s_1)\Big) +(s_{1}\leftrightarrow s_{2})\, .
\en
The final result for the Im$A^{(6)}$ is given by the sum
\eq
\label{eq:sumFSI}
 {\rm Im} A^{(6)}&=& \sum_{i=1,2;\ j=W,\pi}{\rm Im} A^{(6,i)}_{j}+\sum_{j=W,\pi;\
 k=S,P}{\rm Im} A^{(6,3)}_{j,k}\,.
\en

The relation between 
this imaginary amplitude and the functions defined in Appendix 
\ref{Amplitudes} is
\begin{equation}
\im A^{(6)} = \sum _{i=8,27} G_i\,C_i^{(6)}(s_1,s_2,s_3)\,+\,
\sum_{i=1,11}H_i^{(6)}(s_1,s_2,s_3)\widetilde K_i \, .
\end{equation}
This relation is also valid for $K^+\to \pi^0\pi^0\pi^+$.

\subsection{Final State Interactions for  $K^+\rightarrow\pi^0\pi^0\pi^+$}
\label{sec:p006}

The calculation is analogous to the one for 
$K^+\rightarrow\pi^+\pi^+\pi^-$.
The relevant graphs are depicted in Figure \ref{fig:imp6p00}. 
In the case  in which the weak vertex is of ${\cal O}(p^4)$, we get 
\eq
{\rm Im}A^{(6,1)}_W&=& \fr{\s(s_3)}{32\pi}\frac{(s_3- m_\pi^2)}
{f_{\pi}^2}\left[
 2 M_{11}(s_3)+ \widetilde{M}_{11}(s_3)+ \widetilde{M}_{10}(s_3)- 
\widetilde{M}_{12}(s_3) (m_K^2\right. \nonumber \\
&&\left.
+3 m_\pi^2-2
  s_3) + \widetilde{M}_{12}^s(s_3)\right]
\,,\\
{\rm Im}A^{(6,2)}_W&=& \fr{\s(s_3)}{32\pi}\frac{m_\pi^2}{f_{\pi}^2}\left[
  M_{7}(s_3)+\widetilde{M}_{8}(s_3)+\widetilde{M}_{9}(s_3) (m_K^2+3 m_\pi^2-2
  s_3)\right.\nonumber \\
&&
\left. -\widetilde{M}_{9}^s(s_3)\right] \,,
\\
{\rm Im}A^{(6,3)}_{W,S}&=& \fr{\s(s_1)}{64\pi} \frac{(2 m_\pi^2-s_1)}{f_{\pi}^2}\left[
   2 M_{8}(s_1)+\widetilde{M}_{8}(s_1)+\widetilde{M}_{7}(s_1)-\widetilde{M}_{9}(s_1) (m_K^2\right. \nonumber \\ &&+3 m_\pi^2-2
  s_1)
\left.+\widetilde{M}_{9}^s(s_1)\right]
 +(s_1\leftrightarrow s_2)\,.
\en

Also in this case diagram  C generates both 
 $S$-wave and  $P$-wave contributions.
The P-wave contribution due to the diagram  C in \ref{fig:imp6p00} 
 is
\eq
{\rm Im}A^{(6,3)}_{W,P}&=& \fr{\s(s_1)}{64\pi}\frac{1}{f_{\pi}^2} 
\fr{s_1 (s_3-s_2)}{s_1^2-2 (m_K^2+m_\pi^2) s_1+(m_K^2-m_\pi^2)^2}
\left[(s_1-(m_K^2+3 m_\pi^2))\right. \nonumber \\
\et
\times(\widetilde{M}_{8}(s_1)-\widetilde{M}_{7}(s_1)+\widetilde{M}_{9}(s_1)(2
s_1-m_K^2-3  m_\pi^2)) 
+2 \widetilde{M}^s_{8}(s_1) \nonumber \\
\et-2 \widetilde{M}^s_{7}(s_1)+\widetilde{M}^s_{9}(s_1)(5 s_1-3(m_K^2+3
  m_\pi^2))+2 \widetilde{M}^{ss}_{9}(s_1) \nonumber \\
\et
+\fr{8}{3} b^2(s_1) M_{9}(s_1)
  \Big] 
 +(s_1\leftrightarrow s_2) \,.
\en

If the strong  vertex  is ${\cal O}(p^4)$ and the weak
vertex is  order ${\cal O}(p^2)$, we get
\eq
{\rm Im}A^{(6,1)}_\pi&=& \fr{\s(s_3)}{32\pi} 
 \left.(M_{10}(s_1)+M_{10}(s_2))\right|_{p^2}\ 
(P_3(s_3)+\widetilde{P}^{(0,0)}_4(s_3)) \,,
\\
{\rm Im}A^{(6,2)}_\pi&=& \fr{\s(s_3)}{32\pi} \left. (M_{7}(s_3)+M_{8}(s_1)+M_{8}(s_2))\right|_{p^2}\, 
(P_8(s_3)+\widetilde{P}^{(0,0)}_8(s_3)) \,,\nonumber 
\\ \\
{\rm Im}A^{(6,3)}_{\pi ,S}&=& \fr{\s(s_1)}{64\pi}\left.\left(2 M_{8}(s_1)+M_{8}(s_2)+M_{8}(s_3)+M_{7}(s_2)+M_{7}(s_3)\right)\right|_{p^2}\nn\\ \et
\times \Big(\widetilde{P}^{(0,0)}_3(s_1)
+\widehat{P}^{(0)}_{1,4}(s_1)+\widetilde{P}^{(0)}_{2,4}(s_1)\Big)
+(s_{1}\leftrightarrow s_{2}) \,,
\\
{\rm Im}A^{(6,3)}_{\pi ,P}&=&
\fr{\s(s_1)}{64\pi}\left.\left(M_{7}(s_3)-M_{8}(s_3)-M_{7}(s_2)+M_{8}(s_2)\right)\right|_{p^2}\fr{1}{s_1-4 m_\pi^2}\nonumber \\ \et
\times \Big( (s_1-4 m_\pi^2)
(\widetilde{P}^{(0,0)}_3(s_1)-\widehat{P}^{(0)}_{1,4}(s_1)+\widetilde{P}^{(0)}_{2,4}(s_1))+2 \widetilde{P}^{(1,0)}_3(s_1)
\nonumber \\ \et
- 2 \widetilde{P}^{(1)}_{1,4}(s_1) +2\widehat{P}^{(1)}_{2,4}(s_1)\Big)
 +(s_{1}\leftrightarrow s_{2})
\,.
\en
The total contribution is given by the sum  
of (\ref{eq:sumFSI}) with the proper right-hand side terms.

\subsection{Integrals}
\label{sec:integ6}
The integrals necessary to compute the two-bubble 
FSI we discussed in the previous subsection can be calculated 
generalizing the method outlined in \cite{BCEGS97}.
As an example  we show the integration of the function
\eq
\label{eq:Bdefinition}
32 \pi^2 B(m_1,m_2,t)&=& C_B +\left\{\fr{2
    \eta\delta}{t}\ln\fr{\eta-\delta}{\eta+\delta}
+\fr{\la}{t}\ln\fr{(\la-t)^2-\eta^2\de^2}{(\la+t)^2-\eta^2\de^2}  \right\}
\en
where $C_B$ is a  term which does not depend on $t$, 
\eq
C_B &=&
2 \left(1-\ln\fr{\eta^2-\delta^2}{4\nu^2}\right)
\label{eq:cdef}
\en
and
\eq
\eta&=& m_1+m_2\nn\\
\de&=&  m_1-m_2\nn \\
\la&=& \sqrt{ \left[\left( t-\eta^2\right)\left( t-\de^2\right)\right]}\,.
\label{eq:etadelta}
\en
In the center of mass frame one  can define
\eq
Q&=& p_K+p_\pi=(\sqrt{s},0,0,0)\nn \\
p_\pi&=&\fr{\sqrt{s}}{2}\left(\left(1- \fr{m_K^2-m_\pi^2}{s}\right),0,0,
\sqrt{1-\fr{2 (m_K^2+m_\pi^2)}{s}+\fr{(m_K^2-m_\pi^2)^2}{s^2}}\right)\nonumber
\\
\en
where $p_\pi$ is the momentum of the external 
pion entering in the same vertex of the Kaon.
The functions $B$ can also be generated  in the strong vertex.
  In this case $p_k$ is the momentum of an external pion. 
The contribution to the imaginary part of the amplitude $A$ is
\eq
{\rm Im} \, A=\fr{1}{32\pi}\s(s)\int_{-1}^1 
{\rm d} \,\cos\theta\, B[m_1,m_2,t]
\en
with 
\eq
t&=& a+ b \cos\theta
\en
and $a\equiv a(s)$, $b\equiv b(s)$ in (\ref{asbs}).
In order  to solve the difficult part of the integral one can put
\eq
\label{eq:tdef}
t=\fr{1}{2}\left[\eta^2+\de^2-(\eta^2-\de^2)\fr{1+x^2}{2 x}\right]\,.
\en
In this way
\eq
\int_{-1}^1  d\!\cos\theta 
\fr{\la}{t}\ln\fr{(\la-t)^2-\eta^2\de^2}{(\la+t)^2-\eta^2\de^2}&=&
\fr{\eta^2-\delta^2}{2 b}\int_{x_{\rm min}}^{x_{\rm max}} dx\;
  \fr{(1-x^2)^2\ln x}{x^2\left( x^2+1-2 x\al \right)} 
\\
&=& \fr{\eta^2-\delta^2}{2 b x}\biggr\{-1-x^2-(1-x^2)\ln x+\al x \ln^2 x
\nn\\
\et
+2
  x\sqrt{\al^2-1}\biggl(\ln x \ln\fr{1-\al+x\sqrt{a^2-1}}{1-\al-x\sqrt{a^2-1}}
\biggr.
\nn\\
\et
+Li_2\left(\fr{x}{\al+\sqrt{\al^2-1}}\right)\nn\\ \et
\left.
\biggl.
+Li_2\left(x(\al+\sqrt{\al^2-1})\right)\biggr)\biggr\}\right|_{x_{\rm min}}^{x_{\rm max}}
\en
where
\eq
\al&=& {(\eta^2+\delta^2)\over (\eta^2-\delta^2)}\nonumber\\
x_{\rm max}&=&\fr{2}{\eta^2-\delta^2}\left\{
\fr{\eta^2+\delta^2}{2}-a-b+\fr{1}{2}\sqrt{(2
  (a+b)-(\eta^2+\delta^2))^2-(\eta^2-\delta^2)^2}\right\}\nn \\
x_{\rm min}&=&\fr{2}{\eta^2-\delta^2}\left\{
\fr{\eta^2+\delta^2}{2}-a+b+\fr{1}{2}\sqrt{(2
  (a-b)-(\eta^2+\delta^2))^2-(\eta^2-\delta^2)^2}\right\}\,.\nn \\
\label{eq:xdef}
\en
In the case $m_K=m_\pi$, $a+b=0$ and one recovers the 
formulas of \cite{BCEGS97}.

\subsection{Analytical Results for the Dominant FSI Phases at NLO}
\label{phasesNLO}

The elements of the matrices defined in (\ref{Rdelta2def}) have the next 
analytical expressions at NLO
\ba
\mathbb{R}^{\rm LO} &=& \left(\begin{array}{cc}R_{11}&R_{12}
\\R_{21}&R_{22}\end{array}\right)\,,
\nonumber\\
\delta_2^{\rm NLO}&=&\frac{\dis \sum_{i=8,27,E} G_i \left(C_{i,1}^{(++-)}
+C_{i,1}^{(00+)}\right)
+\sum_{i=1,11} \left(H_{i,1}^{(6)(++-)}+ H_{i,1}^{(6)(00+)}\right)
\widetilde K_i }
{\dis \sum_{i=8,27,E} G_i  \left(B_{i,1}^{(++-)}+B_{i,1}^{(00+)}\right)
+\sum_{i=1,11} \left(H_{i,1}^{(4)(++-)}+ H_{i,1}^{(4)(00+)}\right)
\widetilde K_i }\,,
\ea
with 
\ba
R_{11} &=& \frac{\left(-\beta_1+\frac{1}{2}\beta_3\right)^{\rm NR}
\left(\alpha_1+\alpha_3\right)^{\rm R-NR}-
\left(\beta_1+\beta_3\right)^{\rm NR}
\left(-\alpha_1+\frac{1}{2}\alpha_3\right)^{\rm R-NR}}
{\left(-\beta_1+\frac{1}{2}\beta_3\right)^{\rm NR}
\left(\alpha_1+\alpha_3\right)^{\rm NR}-
\left(\beta_1+\beta_3\right)^{\rm NR}
\left(-\alpha_1+\frac{1}{2}\alpha_3\right)^{\rm NR}}
\, ,\nonumber\\
R_{21} &=& \frac{\left(-\beta_1+\frac{1}{2}\beta_3\right)^{\rm NR}
\left(\beta_1+\beta_3\right)^{\rm R-NR}-
\left(\beta_1+\beta_3\right)^{\rm NR}
\left(-\beta_1+\frac{1}{2}\beta_3\right)^{\rm R-NR}}
{\left(-\beta_1+\frac{1}{2}\beta_3\right)^{\rm NR}
\left(\alpha_1+\alpha_3\right)^{\rm NR}-
\left(\beta_1+\beta_3\right)^{\rm NR}
\left(-\alpha_1+\frac{1}{2}\alpha_3\right)^{\rm NR}}
\ ,\nonumber\\
R_{12} &=& -\frac{\left(-\alpha_1+\frac{1}{2}\alpha_3\right)^{\rm NR}
\left(\alpha_1+\alpha_3\right)^{\rm R-NR}-
\left(\alpha_1+\alpha_3\right)^{\rm NR}
\left(-\alpha_1+\frac{1}{2}\alpha_3\right)^{\rm R-NR}}
{\left(-\beta_1+\frac{1}{2}\beta_3\right)^{\rm NR}
\left(\alpha_1+\alpha_3\right)^{\rm NR}-
\left(\beta_1+\beta_3\right)^{\rm NR}
\left(-\alpha_1+\frac{1}{2}\alpha_3\right)^{\rm NR}}
\, ,\nonumber\\
R_{22} &=& -\frac{\left(-\alpha_1+\frac{1}{2}\alpha_3\right)^{\rm NR}
\left(\beta_1+\beta_3\right)^{\rm R-NR}-
\left(\alpha_1+\alpha_3\right)^{\rm NR}
\left(-\beta_1+\frac{1}{2}\beta_3\right)^{\rm R-NR}}
{\left(-\beta_1+\frac{1}{2}\beta_3\right)^{\rm NR}
\left(\alpha_1+\alpha_3\right)^{\rm NR}-
\left(\beta_1+\beta_3\right)^{\rm NR}
\left(-\alpha_1+\frac{1}{2}\alpha_3\right)^{\rm NR}}\,,
\ea
The definitions of $\alpha_1$, $\alpha_3$, $\beta_1$ and $\beta_3$ are in 
(\ref{amp1}) and the values of their relevant combinations are
\ba
\left(-\alpha_1+\frac{1}{2}\alpha_3\right)^{\rm NR} &=&
\sum_{i=8,27,E} G_i B_{i,0}^{(00+)}
+\sum_{i=1,11} H_{i,0}^{(4)(00+)} \widetilde K_i \, ,\nonumber\\
\left(-\alpha_1+\frac{1}{2}\alpha_3\right)^{\rm R-NR} &=&
\sum_{i=8,27,E} G_i C_{i,0}^{(00+)}
+\sum_{i=1,11} H_{i,0}^{(6)(00+)} \widetilde K_i\, ,\nonumber\\
\left(-\beta_1+\frac{1}{2}\beta_3\right)^{\rm NR} &=& \frac{1}{2}
\left\lbrack \sum_{i=8,27,E} G_i \left(B_{i,1}^{(++-)}-B_{i,1}^{(00+)}
\right)\right.\nonumber\\
&&\left.+\sum_{i=1,11} \left(H_{i,1}^{(4)(++-)}- H_{i,1}^{(4)(00+)}\right)
\widetilde K_i 
\right\rbrack\, ,\nonumber\\
\left(-\beta_1+\frac{1}{2}\beta_3\right)^{\rm R-NR} &=& \frac{1}{2}
\left\lbrack \sum_{i=8,27,E} G_i \left(C_{i,1}^{(++-)}-
C_{i,1}^{(00+)}\right)\right.\nonumber\\
&&\left.+\sum_{i=1,11} \left(H_{i,1}^{(6)(++-)}-H_{i,1}^{(6)(00+)}\right)
 \widetilde K_i \right\rbrack\, ,\nonumber\\
\left(\alpha_1+\alpha_3\right)^{\rm NR} &=&
\sum_{i=8,27,E} G_i B_{i,0}^{(+-0)}
+\sum_{i=1,11} H_{i,0}^{(4)(+-0)} \widetilde K_i \, ,\nonumber\\
\left(\alpha_1+\alpha_3\right)^{\rm R-NR} &=&
\sum_{i=8,27,E} G_i C_{i,0}^{(+-0)}
+\sum_{i=1,11} H_{i,0}^{(6)(+-0)} \widetilde K_i \, ,\nonumber\\
\left(\beta_1+\beta_3\right)^{\rm NR} &=& -\frac{1}{2}\left\lbrack
\sum_{i=8,27,E} G_i B_{i,1}^{(+-0)}
+\sum_{i=1,11} H_{i,1}^{(4)(+-0)} \widetilde K_i\right\rbrack
\, ,\nonumber\\
\left(\beta_1+\beta_3\right)^{\rm R-NR} &=& \frac{1}{2}\left\lbrack
\sum_{i=8,27,E} G_i C_{i,1}^{(+-0)}
+\sum_{i=1,11} H_{i,1}^{(6)(+-0)} \widetilde K_i\right\rbrack\, .
\ea
where the functions $B_{i,0(1)}$, $C_{i,0(1)}$ and  $H_{i,0(1)}$  
are those obtained form the expansion in  (\ref{yexpansion})
of the corresponding full quantities that can be found
in Appendix \ref{ANLO}.

Disregarding the tiny CP-violating (less than 1\%)
and the effects of order $e^2p^2$ (the loop contribution
 is less than 2\%), 
we obtain the numbers in (\ref{RmatrixNLO}).

\end{document}